# Soft magnetic microrobot doped with porous silica for stability-enhanced multimodal locomotion in nonideal environment

*Shangsong Li#, Dong Liu#, Yuping Hu#, Zhijie Su, Xinai Zhang, Ruirui Guo, Dan Li, Yuan Lu\**

*Shangsong Li, Dong Liu, Yuping Hu, Zhijie Su, Xinai Zhang, Ruirui Guo, Dan Li, Yuan Lu*
Key Laboratory of Industrial Biocatalysis, Ministry of Education
Department of Chemical Engineering
Tsinghua University
Beijing 100084, China

*Shangsong Li*
School of Materials Science and Engineering
Tsinghua University
Beijing 100084, China

*Xinai Zhang*
Beijing No.4 High School International Campus
Beijing 100034, China

*Ruirui Guo*
College of Bioengineering
Tianjin University of Science and Technology
Tianjin 300457, China

#: These authors contributed equally to this work.

*: Corresponding author; Email: yuanlu@tsinghua.edu.cn

**Abstract**

As an emerging field of robotics, magnetic-field-controlled soft microrobot has broad application prospects for its flexibility, locomotion diversity as well as remote controllability. Magnetic soft microrobots can perform multimodal locomotion under the control of a magnetic field, which may have potential applications in precision medicine. However, previous researches mainly focus on new locomotion in a relatively ideal environment, lacking exploration on the ability of magnetic microrobot locomotion to resist external disturbances and proceed in a nonideal environment. Here, a porous silica-doped soft magnetic microrobot is constructed for enhanced stability of multimodal locomotion in the nonideal biological environment. Porous silica spheres are doped into NdFeB-silicone elastomer base, improving adhesion properties as well as refining the comprehensive mechanical properties of the microrobot. Multimodal locomotions are achieved, and the influence of porous silica doping on the stability of each locomotion in nonideal environment is explored in depth. Motions in nonideal circumstances such as climbing, loading, current rushing, wind blowing, and obstacle hindering are conducted successfully with porous silica doping. Such a stability-enhanced multimodal locomotion system can be used in biocatalysis as well as thrombus removal, and its prospect for precision medicine is highlighted by *in vivo* demonstration of multimodal locomotion with nonideal disturbance.

# 1. Introduction

Due to outstanding flexibility, locomotion diversity as well as remote controllability, magnetic soft microrobot has broad application prospects. Under the control of an external magnetic field, a micro magnetic soft robot can perform multimodal locomotion, including basic movement patterns such as walking, jumping, rolling, and swimming. This brings a theoretical possibility to the application of magnetic soft microrobot in various areas. Especially, microrobots have potential applications in precision medicine [1], including minimally invasive surgery [2], drug delivery [3], and cancer treatment [4], contributing to human health. The advantages of soft magnetic microrobot are obvious. Firstly, microrobots are usually limited to millimeters or even micrometers. This miniaturization makes it easy to enter certain parts of a complex structure, especially narrow ones such as small intestine and microchemical tubes, reducing the complexity of the entry process [5]. Due to the small size, its movement will have high flexibility and experience little hindrance, not causing damage to surroundings. Secondly, since the control of micro robot movement is realized by the magnetic field, which can easily penetrate the blockings without causing additional changes, the safety and effectiveness of the method are ensured [6]. By changing the magnitude and direction of the magnetic field, various locomotion of the robot can be achieved, and suitable locomotion can be used in different environments to accomplish the whole task [7].

For the research of magnetic soft robot, a lot of attention has been paid to creating new locomotions, such as jellyfish-like swimming [8], helical propulsion [9], flapping [10], and rotating [11]. These studies provided innovation in the basic level of locomotion and broadened the application possibility of magnetic microrobots. However, a common problem is that although many locomotions could be achieved for excellent display effect, the locomotions involved in previous studies were basically completed in a relatively ideal environment, such as walking on a flat surface, rolling on a normal substrate, and swimming in still water. In fact, the stability of locomotion in the nonideal environment is greatly important, because in real internal environment, the moving of magnetically controlled flexible robot will be affected by various disturbances, including but not limited to muscle peristalsis, airflow impact, liquid flow impact, obstacle barrier, sudden change of physical properties of the base, steep slope, and sharp turn. A robot walking normally in an ideal environment may slip or even fall under the impact of water flow. Regarding the ability of magnetic microrobot locomotion to resist external disturbances and proceed in nonideal environment, unfortunately, there is a lack of relevant studies, which is unfavorable to the practical application of micro magnetron robots in precision medicine.

In order to solve the aforementioned problem of locomotion stability in the nonideal environment, two options could be considered. The first option is to adjust external magnetic field according to the change of environment, keeping the robot in a relatively stable situation. However, this idea is difficult to realize in practice, because the microrobot needs to face the complex and changeable actual environment in various tasks. With the deep reinforcement learning method, it is still difficult to give an appropriate and timely response to various environmental changes in microscale biological environments [12]. The second option is to control the performance of the robot body material to enhance its adhesion to the substrate and improve its mechanical properties. In this way, the motion stability of the microrobot can be improved greatly. In fact, material modification by doping is a common method to improve material properties and prepare new materials [13]. For the improvement of adhesion properties, non-metallic inclusions with high porosity and rough micro surface can be introduced, such as porous silica rods or spheres [14,15]. The regulation of the matrix polymer itself may also improve adhesion [16]. However, the modification of soft materials often pointed to applications such as flexible electronics and percutaneous drug delivery [14,17]. In previous studies, unfortunately, there is no research directly combining material doping modification with multimodal locomotion of soft magnetic microrobot.

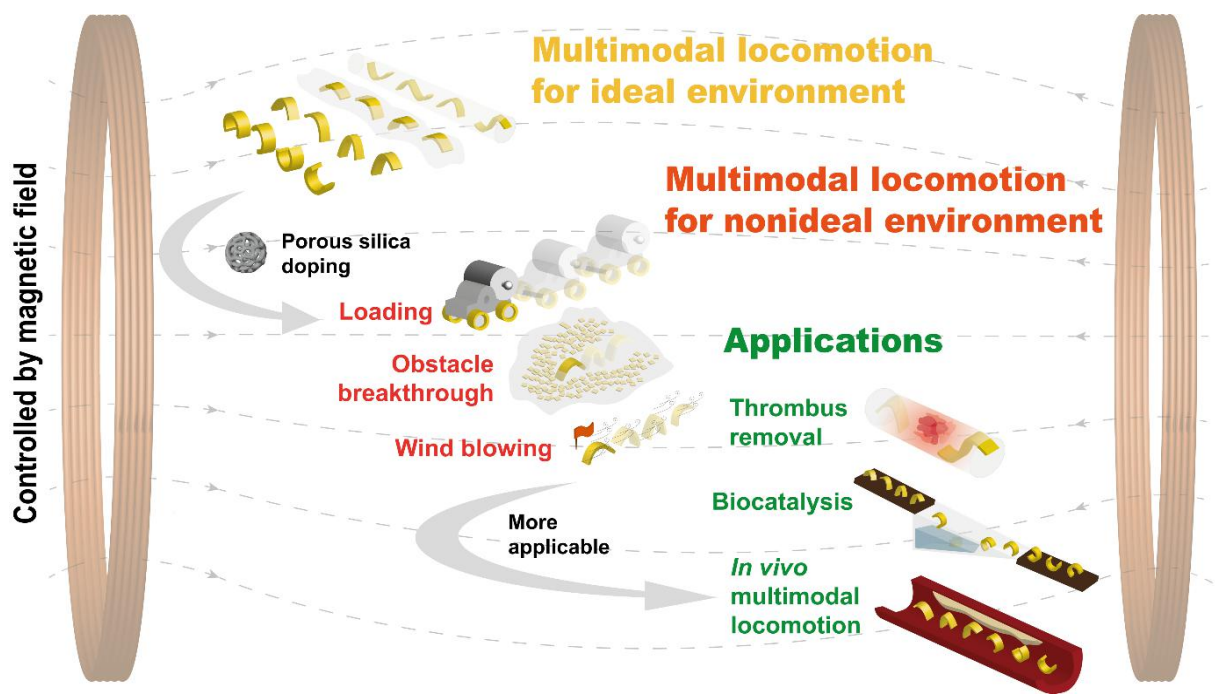

**Fig. 1.** An overall illustration of the porous silica-doped soft magnetic microrobot for enhanced stability of multimodal locomotion in nonideal biological environment.

Here, a soft magnetic microrobot with porous silica doping was first reported for stability-enhanced multimodal locomotion in the nonideal environment (Fig. 1). Porous silica spheres were doped into an NdFeB-silicone elastomer base, improving adhesion properties as well as refining the comprehensive mechanical properties of the magnetic soft microrobot. Basic multimodal locomotions including rolling, walking, jumping as well as crawling, were achieved, and the influence of porous silica doping on the stability of each locomotion in the nonideal environment was explored in depth. With the stability data of each locomotion, basic locomotions could be assembled into more complex locomotions (drifting, stirring, trolley, and boat) via rational design, enlarging the locomotion library while achieving more complex tasks. Different from previous action demonstrations in the ideal environment, among the locomotion library of this research, motions could be conducted in nonideal circumstances such as climbing, loading, current rushing, wind blowing, and obstacle hindering. Such stability-enhanced multimodal locomotion system could be used in biocatalysis as well as thrombus removal, and its prospect for precision medicine was highlighted by *in vivo* demonstration of multimodal locomotion with nonideal disturbance.

## 2. Material and methods

### 2.1 Chemicals

Triethanolamine (TEA), hexadecyl trimethyl ammonium bromide (CTAB), sodium salicylate (NaSal), tetraethyl orthosilicate (TEOS), diammonium ABTS, glucose, glucose oxidase (GOx), horseradish peroxidase (HRP), NdFeB magnetic microparticles, oleic acid, ethylene glycol, and ethanol were purchased from Sigma-Aldrich. Ecoflex 00-10 polymer matrix was purchased from Smooth-On Inc. Silicone rubber, polyurethane (PU), and polytetrafluoroethylene (PTFE) were purchased from DuPont Chemical Co., Ltd. Deionized water was purified using a Milli-Q system. All other chemicals were used as received.

### 2.2 Synthesis of Dendritic Mesoporous Silicon Nanoparticles

60 μL TEA was added to 25 mL deionized water, and the mixture was shaken at 160 rpm in a water bath at 80 °C for 0.5 h. Then 380 mg CTAB and 168 mg NaSal were added and continued shaking was applied for 1 h. Afterward, 4 mL TEOS was added, and a continued shaking was applied for 2 h. Then the mixture was centrifuged at 7000 rpm for 20 min to collect the products. The products were washed with ethanol for three times to remove the residual reactants, and finally dried overnight in a vacuum at room temperature.

### 2.3 Preparation of Magnetic Soft Microrobot

The composition of the robot involved neodymium iron boron (NdFeB) magnetic micro-particles (MQP 15-7, Magnequench; average diameter: 5 μm, density: 7.61 g·cm$^{-3}$), the Ecoflex 00-10 polymer matrix A and B (Smooth-On Inc.; density:1.04 g·cm$^{-3}$), and porous silica (p-Si) microparticles. The amount of NdFeB, Ecoflex A, and Ecoflex B added for each glass slab was determined in correspondence to the thickness of the layer and area of the slab, and the amount of porous silica added was calculated through the demanded concentration of the material.

Two specifications were made for different uses throughout the experiment. The first specification created films of 1mm. Slabs of approximately 100 cm$^2$ were used as a model for film construction. Ecoflex A (4.4 g) was added onto the glass slab first, and then NdFeB micro-particles (8.8 g) were added onto the polymer matrix Ecoflex A on slab in order to avoid dispersion to other places. Porous silica microparticles were added in the calculation of concentration (for example, the preset 0.2% concentration involved 0.0352 g of porous silicon micro-particles). Ecoflex B (4.4 g) was added at last, since it started to react with Ecoflex A to form crosslinking structure once added. The second specification created films of 0.2 mm, which involved glass slab of 48 cm$^2$ and the addition of 1 g NdFeB micro-particles, 0.5 g Ecoflex A, 0.5 g Ecoflex B, and porous silica particles in correspondence to concentration.

After the addition of materials, the stirring glass rod was used to fully mix the mixture on the glass slab and to ensure that film of uniform thickness was made. The uniformly mixed mixture was placed in the air for 5 hours to solidify.

## 2.4 Measurements and Characterization

SEM images were recorded on a Zeiss Gemini high-resolution scanning electron microscope (Germany) operating at 5 kV. The EDS images were recorded while operating at 15 kV. Microrobot samples were first put into liquid nitrogen and then broke, leaving quenched sections for observation. TEM images were taken with a JEOL 2011 microscope (Japan) operated at 200 kV. Magnetic characterization was conducted on a vibrating sample magnetometer LakeShore 8604.

## 2.5 Magnetization of Magnetic Soft Microrobot

After solidification, the film was cut into pieces, becoming the robot bodies. The cutting involved two specifications, each corresponding to the specific thickness of the film. For films of 1 mm (h=1 mm), each robot had dimensions of w=3.6 mm and L=8.9 mm; for films of 0.2 mm (h=0.2 mm), each robot had dimensions of w=1.5 mm and L=3.7 mm.

The robot body of 8.9 mm length was then wrapped around a cylindrical wood rod with a circumference of 8.9 mm; the robot with 3.7 mm length was wrapped around a rod with 3.7 mm circumference. In order to receive uniform magnetization, the robots were vertically wrapped around the rod without any overlapping. The soft-bodied robots were then put into the pulse magnetizer (Beijing EUSCI Technology Limited), which could give a magnetic pulse of 2.5 T for magnetization. The angles by which robots were put into the magnetizer turned to be important, as it necessarily created a phase shift *β* in the magnetization profile ***m***. It was determined by the angle between the vertical direction and the wrapping start point. After magnetization, the robot was unwrapped, and a harmonic magnetization profile was achieved.

## 2.6 Tensile Test and Tensile-Adhesion Test

Tensile test of materials containing different concentrations of porous silica was performed to directly reveal the mechanical properties of the synthesized material. The tensile test was performed at room temperature on a universal testing machine (SHIMADZU) with 50 N load cells. Samples of rectangular shape were used with the width of 10 mm, length of 40 mm, and height of 1 mm were used. The samples were placed between two platforms, with both ends gripped by the machine grips: upper in the moving grips and lower in the fixed grips. The rate of stretching was constantly 60 mm·min$^{-1}$. The mechanical parameters were calculated from the stress-strain curve.

With a similar method, properties of adhesion and toughness were also measured, which was called as the tensile-adhesion test. The tensile-adhesion test was performed at room temperature on a universal testing machine (SHIMADZU) with 50 N load cells. To measure the property of adhesion, samples with the size of 37 mm in width, 38 mm in length, and 1 mm in height were used. The samples were pre-settled onto a flat horizontal square platform. The platform had a vertical handle for the testing machine to grip. As a whole, the device was T-shaped. The samples were placed between two platforms, with both handles gripped by the machine grips: upper in the moving grips and lower in the fixed grips. The rate of stretching was constantly 60 mm/min. The related parameters were calculated from the stress-strain curve.

**2.7 Realization of Multimodal Locomotion**

Details were listed in the supplementary information.

**3. Results and Discussion**

**3.1 Microrobot Fabrication**

**3.1.1 Material Synthesis and Characterization**

The material of magnetic microrobot, including both base material and doping material, was synthesized and characterized. For the base material of the robot body, neodymium iron boron (NdFeB) magnetic micro-particles were mixed with Ecoflex 00-10 polymer matrix A and B (a kind of silicone elastomer) according to the mass ratio of 1:1 [1]. NdFeB microparticles provided the ability for magnetic control, while Ecoflex 00-10 polymer matrix made up the viscoelastic and biocompatible body (Fig. 2a). When the raw materials were initially mixed, the mixture was liquid. After standing in the air for 3-5 hours, monomers were crosslinked, forming a silicone network. Eventually, the mixture solidified into silicone elastomer and was made into pieces of microrobots (Fig. 2b). Porous silica (p-Si) microparticles were synthesized via a traditional method and doped into the robot body with a concentration gradient of 0 wt%, 0.3 wt%, 0.6 wt%, 0.9 wt%, and 1.2 wt% [18]. Because of the porous and loose surface (Fig. 2c-d), porous silica may effectively hinder the movement of molecular chains in the viscoelastic matrix, so as to improve the viscosity and refine the mechanical properties of the robot. Magnetic response measurements using a vibrating sample magnetometer (Fig. 2h) showed the classical hysteresis loop of hard magnetic materials, as well as gave the magnetization saturation (MS) value of 109.4 emu/g for NdFeB and 56.9 emu/g for the magnetic soft robot, indicating strong response to the magnetic field. From the macro scale, NdFeB was uniformly mixed in the silicone matrix. However, from the micro scale, obvious segregation was observed via SEM picture of the quenched section (Fig. 2e). Element mapping of Si, Fe, and O elements further confirmed the granular distribution of NdFeB microparticles (Fig. 2f-g).

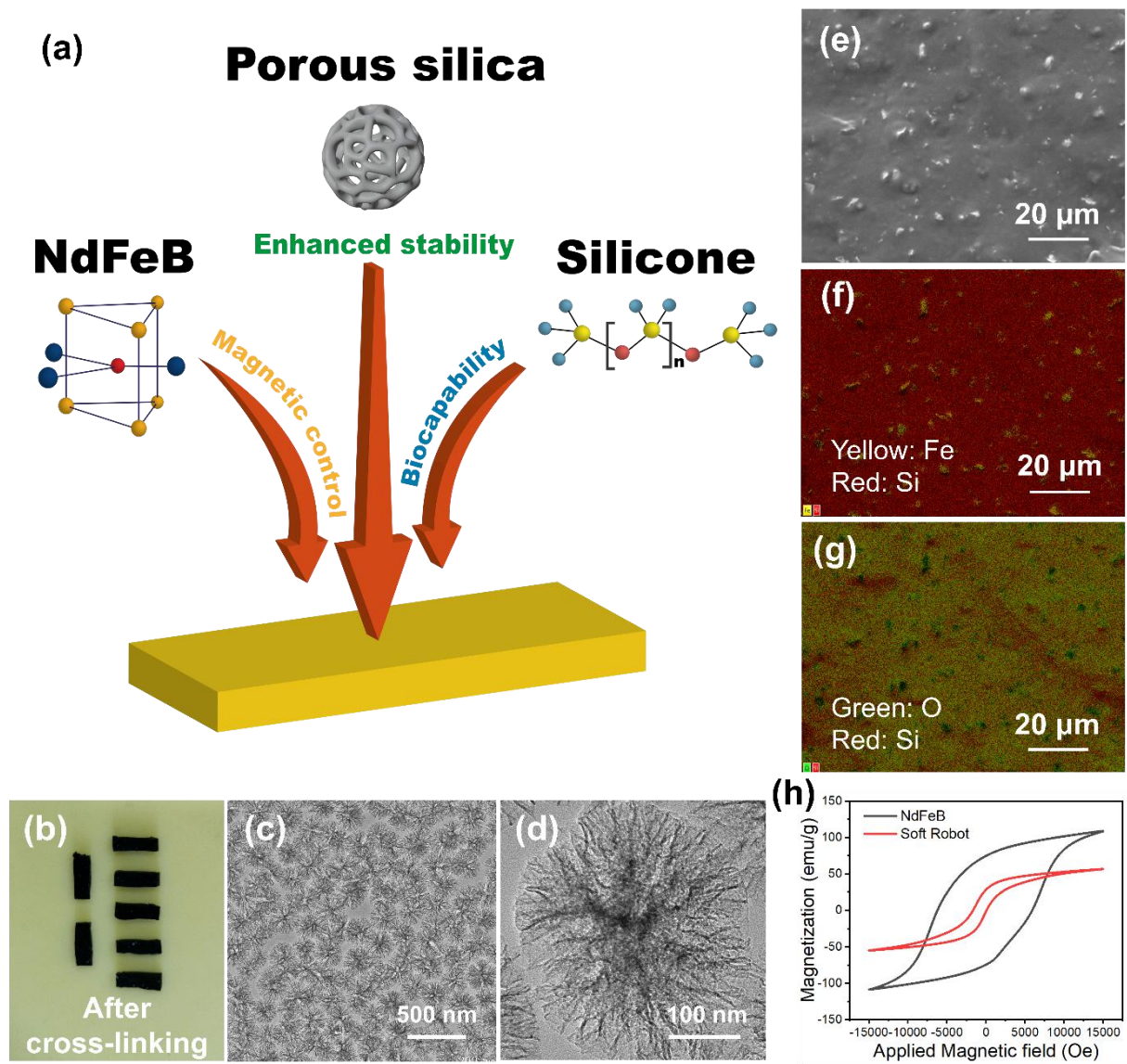

**Fig. 2.** Material synthesis and characterization. (a) An illustration of materials composing the microrobot. (b) The solid mixture cut into pieces of microrobot after crosslinking. (c) Transmission Electron Microscope (TEM) image of porous silica nanoparticle, low-magnification. (d) TEM image of the porous silica nanoparticle, high-magnification. (e) Scanning Electron Microscope (SEM) image for the quenched section of the magnetic soft robot. (f) Energy Dispersive Spectrometer (EDS) image of the magnetic soft robot, showing the distribution of the Fe (yellow) and Si (red) elements. (g) EDS image of the magnetic soft robot, showing the distribution of the O (green) and Si (red) elements. (h) Magnetic hysteresis loops of NdFeB and magnetic soft robot.

### 3.1.2 Mechanical Properties

*3.1.2.1 Tensile test*

Mechanical properties of silicone elastomers were investigated via tensile test (Fig. 3b, Video S1) and tensile-adhesion test (Fig. 3a, Fig. 3c). Four groups of tensile properties, elongation, toughness, tensile strength, and elastic modulus were calculated according to the stress-strain curves (Fig. 3d). When the concentration of doped porous silica increased from zero, the elongation (Fig. 3e) and toughness (Fig. 3f) of silicone elastomer showed an obvious upward trend, maximized by the porous silica incorporation of 0.9 wt%, and then declined for 1.2 wt%. The results could be explained intuitively that, after being embedded with different concentrations of porous silica (from 0 to 1.2 wt%), silicone elastomers became softer and easier to extend, and was harder to break when stretching slowly. To a certain extent, it avoided the soft robot from being broken by some small but continuous tension. Meanwhile, the elastic modulus (Fig. 3h) and tensile strength (Fig. 3g) decreased with the increase of porous silica concentration. The decline of elastic modulus was relatively significant. Decreasing elastic modulus indicated that the soft robot was easier to produce elastic deformation under stress. Summarily, the porous silica doping could refine the mechanical properties by softening the robot body with the best softening effect at 0.9 wt%, so it was feasible to dope porous silica in the silicone elastomer.

*3.1.2.2 Tensile-adhesion test*

For the enhancement of locomotion stability, more adhesion to substrate was required. Porous silica doping was a method to improve adhesion via rational design. The enhanced adhesiveness of silicone elastomer through the incorporation of porous silica was investigated by a tensile-adhesion test [15]. Silicone elastomers embedded with different concentrations of porous silica (from 0 to 1.2 wt%) were freshly prepared. Thereafter, the silicone elastomer was attached to different substrates, and the adhesiveness was measured through the stress-strain curve (Fig. 3i, Fig. S4). Three materials of substrate were selected in the test, including biomimetic silicone rubber closing to skin texture (hereinafter referred to as silicone), medical polyurethane (hereinafter referred to as PU), and medical polytetrafluoroethylene (hereinafter referred to as PTFE). Such substrate selection was meaningful for exploring the application prospects of microrobots in biomedicine. Subsequently, the differences of microrobot locomotions with gradient concentrations on these substrates were also discussed to explore whether the results corresponding to the measurements of mechanical properties of these three substrates. Biomimetic silicone rubber could work as a good simulation for skin texture.[19] PU has good wear resistance, toughness, excellent corrosion resistance and chemical resistance, soft tissue compatibility and blood compatibility. It could be used as medical hose, bone repair material, supplementary material for vascular surgical suture, and the main constituent material of artificial heart and artificial kidney [20]. PTFE is the main material of artificial blood vessel [21]. Therefore, PU and PTFE could well simulate the human internal environment for microrobot locomotion. With silicone elastomer attached to three different substrates, the adhesive strength increased by increasing the concentration of incorporated porous silica (Fig. 3k), and the adhesive energy increased more significantly (Fig. 3j). The maximum adhesive energy on PU substrate, 7.3 J·m$^{-2}$, appeared at 1.2 wt% porous silica concentration, which was approximately 2.5 times higher than that of 0 wt%. The tensile-adhesion test indicated that porous silica doping evidently increased the adhesiveness of the silicone elastomer.

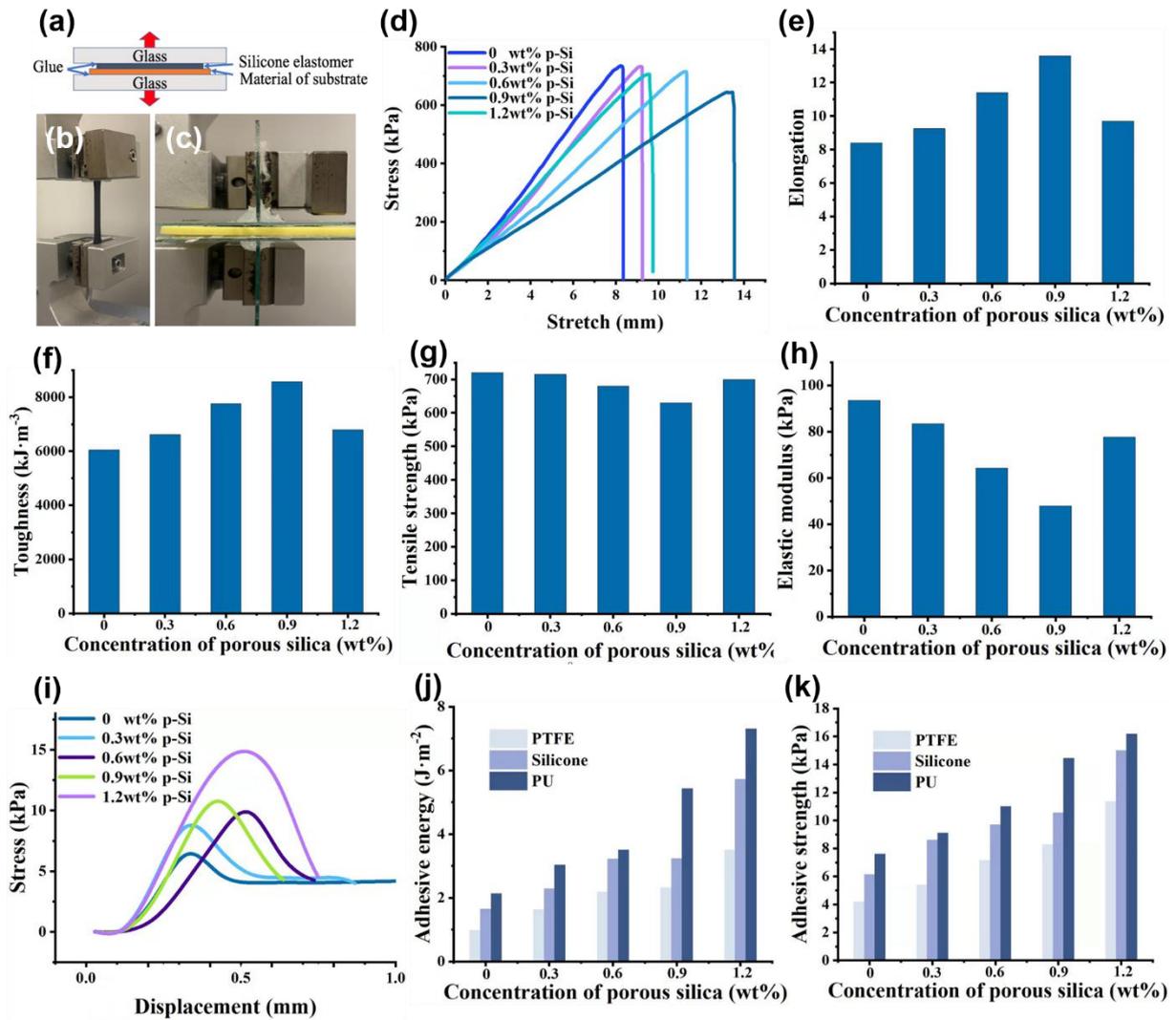

**Fig. 3.** Characterization of mechanical properties of the microrobot material. (a) An illustration of the tensile-adhesion test. (b) A capture of the tensile test. (c) A capture of the tensile-adhesion test. (d) A stress-strain curve for the tensile test. (e) A bar graph showing the relationship between elongation and porous silica concentration. (f) A bar graph showing the relationship between toughness and porous silica concentration. (g) A bar graph showing the relationship between tensile strength and porous silica concentration. (h) A bar graph showing the relationship between elastic modulus and porous silica concentration. (i) A stress-strain curve for the tensile-adhesion test on PU substrate. (j) A bar graph showing the adhesive energy for microrobot material with different concentration of porous silica doping on three different substrates (PTFE, silicone, and PU). (k) A bar graph showing the adhesive strength for microrobot material with different concentration of porous silica doping on three different substrates (PTFE, silicone, and PU).

## 3.2 Multimodal Locomotion

### 3.2.1 Flexible Micro Magnetron Robot for Multimodal Locomotion

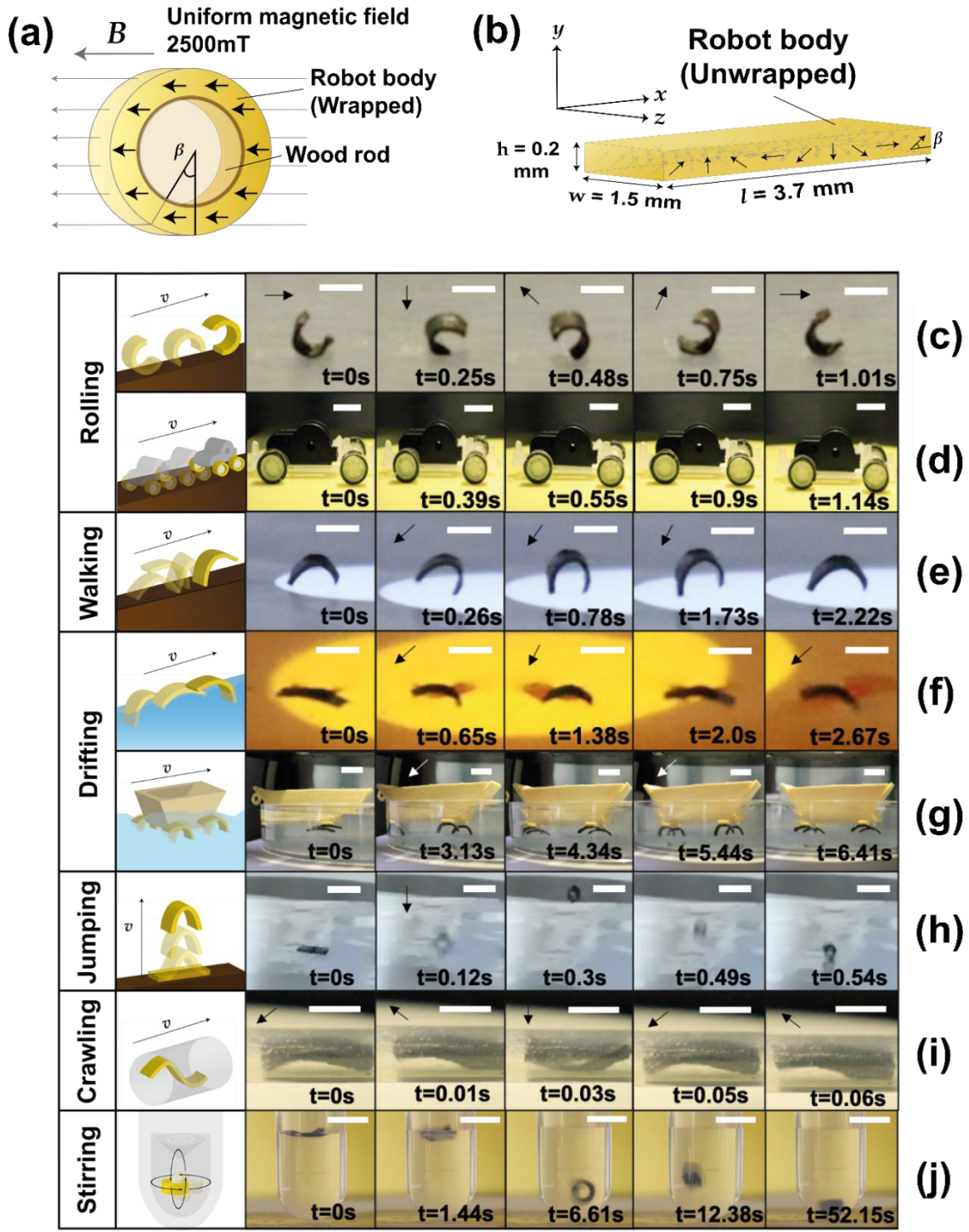

**Fig. 4.** Schematic of magnetization and different locomotions. (a) The robot is wrapped by a wood rod when magnetizing. (b) The robot is unwrapped and showing the magnetization profile *m*. (c) Rolling

locomotion, with a rotating magnetic field $B_{xz}$. Scale bar, 2 mm. (d) A trolley assembled by four rolling robots. Scale bar, 1 cm. (e) Walking locomotion, induced by a three-phase magnetic field. Scale bar, 2 mm. (f) Drifting locomotion, induced by a three-phase magnetic field. Scale bar, 2 mm. (g) A boat assembled by four drifting robots. Scale bar, 2 mm. (h) Jumping locomotion, induced by a large deformation. Scale bar, 5 mm. (i) Crawling locomotion with a rotating magnetic field. Scale bar, 2 mm. (j) Stirring locomotion, with three orthogonal axes of rotating magnetic field. Scale bar, 8 mm.

To achieve various locomotions, a harmonic magnetization profile $m$ was embedded in the robot body. The robot was wrapped with a cylindrical wood rod, whose circumference was 3.7 mm. Meanwhile, a proper phase shift $β$ was significantly important.[1] The bottom was defined as 0°, and via calculation, the default phase shift $β$ was set to 45° in the following research unless specially clarified. The properly wrapped robot was placed into a pulse magnetizer for magnetization with a strong, directional, uniform magnetic field $B$ of 2.5 T (Fig. 4a, Fig. S5).

The dispersed NdFeB microparticles were magnetized into the same direction. Hence, after unwrapping, a robot body with harmonic magnetization profile $m$ would be achieved, and the dimension of microrobot was 0.2×1.5×3.7 mm (Fig. 4b). Owing to these features, different locomotions could be achieved by applying an alternating magnetic field $B = [B_x, B_y, B_z]^T$ in an electromagnetic setup, which contained three pairs of orthogonal Helmholtz coils (Fig. S6). By adjusting the voltage applied on each coil, the magnetic field intensity could be controlled properly to actuate the magnetic microrobot.

An overall description of multimodal locomotion was shown (Fig. 4c-j), which illustrated time-depending captures of each locomotion. In addition, the general theory and mathematical analysis were appended in the supplementary information (Text S1). In the rolling locomotion (Fig. 4c, Video S2), a rotating magnetic field $B_{xz}$ ($f$ = 3 Hz, $B$ = 8 mT) was applied. Compared to other locomotions, the rolling locomotion was an efficient and stable way to move (Fig. 4c). Moreover, microrobots were capable of overcoming rough terrains via rolling locomotion, especially moving on bionic tissue which would be further discussed later. For more complex locomotion, four robots were rationally designed as the four wheels of a magnetically controlled trolley (Fig. 4d, Video S3). The trolley could move smoothly on most substrates and even deliver cargoes.

Walking locomotion was inspired by inchworms (Fig. 4e, Video S4). By applying a periodic magnetic field $\boldsymbol{B}_{xy}$, which presented an increase-invariant-decrease trend in magnitude and invariant-decrease-invariant in a angle of the direction $α$ (maximum $B_{max}$ = 10 mT, minimum $B_{min}$ = 0 mT, maximum $α_{max}$ = 150° and minimum $α_{min}$ = 120°). Moreover, mesoporous silica doping played a positive role in walking locomotion by reinforcing friction.

For the movement in the actual biological environment, microrobots should not only move on land, but also be able to move in liquid. Microrobot in this research was made of a hydrophobic material; consequently, it could float on liquid, especially liquid with high surface tension. By applying the same magnetic field as walking locomotion on a floating robot, drifting locomotion was achieved (Fig. 4f, Video S5). Via this locomotion, microrobot could move quickly, which was coincidentally familiar to Gerridae, and there were a few applications *in vivo* like drifting in blood vessels, GI tract, etc. Similar to the trolley mentioned above, four robots were rationally designed to assemble into a tiny magnetically controlled boat (Fig. 4g, Video S5), which might provide an excellent outlet for drug delivery *in vivo*.

Jumping locomotion could be classified into two ways (Fig. 4h, Video S6), vertical jumping and horizontal jumping. For vertical jumping, the robot was magnetized with phase shift $β$ = -90°, and a magnetic field controlled by DC power with $α$ = 270° (perpendicular to the ground) should be applied. For horizontal jumping, the phase shift $β$ was 0°, and direction of applied magnetic field was same as that of vertical jumping. The magnitude of the magnetic field in both kinds of jumping was large in order to actuate enough deformation at the initial state. Via the combination of vertical and horizontal jumping, microrobots could effectively jump over obstacles, achieving more places not to be reached by rolling or walking.

Crawling locomotion could be achieved by applying an alternative magnetic field $B_{xz}$ with $f$ = 15 Hz and $B$ = 10 mT (Fig. 4i, Video S7). By placing the robot in a glass tunnel with a radius $r$ = 0.6 mm, the robot would crawl like a traveling wave with confined phase transfer actuated by rotating magnetic field. Crawling locomotion might contribute to invasive surgery for its ability to remove thrombus in blood vessels.

Stirring locomotion, with the fundamental of rolling locomotion, was controlled by a magnetic field rotating periodically in three consecutive planes (xy, xz, yz plane). As a result, it enabled the microrobot to stir adequately in order to increase the reaction rate in a small reactor (Fig. 4j, Video S8). Stirring locomotion is promising for intelligent microscale chemical engineering systems by replacing traditional magnetic stirrers, and might also be used for targeted therapy via oriented biocatalysis.

Although there were various locomotions to achieve, the porous silica doping could always enhance locomotion stability, making microrobots more adaptive in the nonideal environment. For rolling and walking locomotions, the porous silica doping could directly increase the friction between microrobot body and substrate, thus making the locomotion more stable. For other locomotions, the effect of the porous silica doping induced stability was mainly because porous silica could soften the microrobot body and increase flexibility. The enhanced flexibility reduced the effect of disturbance from the surroundings, making microrobots smoother to move forward.

### 3.2.2 Rolling Locomotion and Magnetically controlled Trolley

The magnetization orientation of the microrobot was a trigonometric function expanded by a circle; then the microrobot might return to the arc shape and go forward by applying an appropriate sine-cosine varying magnetic field. This led to the microrobot being most suitable for rolling dynamics in essence, and other locomotions were derived from rolling locomotion by tumbling of certain parts in a more complex magnetic field. By means of rolling, the microrobots could achieve maximum moving efficiency due to little friction losses and easy acceleration [22]. Thus, rolling became the major locomotion for microrobots in this study.

The soft robot could roll directionally in a C-shape configuration (Fig. 4c, Video S2), enabled by a rotating magnetic field, which was the rolling locomotion. Similarly, the robot was connected end to end to form a circular shape which could further become part of a magnetically controlled trolley. Doped porous silica could effectively improve rolling locomotion under the circumstances of relatively small friction or large resistance, which was possibly attributed to enhanced adhesion. According to basic mechanical analysis (Fig. 5a), while rolling individually or as a magnetically controlled trolley, microrobot locomotion was mainly affected by three forces: magnetic force, frictional force (represented as $f$ in Fig. 5a), and adhesive force (represented as $f_{ax}$ and $f_{ay}$ in Fig. 5a). Magnetic force enabled the robot to rotate *in situ*, and frictional force made it possible for the robot to move forward (Text S2). The essence of adhesive force was intermolecular attraction [23]. During the macroscopic rolling locomotion, adhesive force could be decomposed into two orthogonal component forces. One was along the direction of motion, and the other was perpendicular to the direction of motion, pointing to the contact surface. The former component had the same effect as friction, that is, the driving force to let the robot move forward. Due to the difference in the order of magnitude, the latter component could be reasonably ignored in the analysis. According to experimental results (Fig. 3), doped porous silica could effectively improve the adhesion of microrobots to specific materials (PTFE, PU, and silicone). This helped with rolling locomotion under a small frictional force or large resistance. It was natural to consider these circumstances, as there were many surfaces of the human body with a low friction coefficient, such as sweat-soaked skins, mucus-covered organ walls, and lipid-rich organs.

Additionally, the rolling locomotion went well under a wide range of friction and resistance, which further broadened the applicability of microrobots in complex human environments. In this study, three groups of experiments were conducted to verify the above theory.

First, oleic acid was added to the surface of a weighing paper to construct a substrate with relatively little friction for rolling locomotion. Robots rolled at different speeds with different concentrations of doped porous silica (Fig. 5b). Those with higher concentrations of porous silica had a larger adhesive force, which offset the decrease of frictional force due to oiled paper substrate and effectively avoided slipping, thus increasing the rolling speed.

Second, set up resistance with a slope. A simple slope was built to change the effect of gravity on robots' rolling locomotion, and the slope structure could be an approximation for some steep motion paths of microrobots in human body. According to mechanical analysis of the robot rolling forward on slope (Fig. 5c), gravity (represented as *mg* in Fig. 5c) could be decomposed into two orthogonal components. The gravity component in the vertical motion direction for slope surface was smaller than the whole gravity *mg* on the flat surface, which reduced the friction. More importantly, the component parallel to the motion direction greatly resisted the rolling locomotion. Considering the above factors, the adhesive force became even more prominent. In the experiment, three different materials were used as slope substrates. Material type of substrate and concentration of doped porous silica were treated as two variables, and the distance between the highest point that microrobots rolled to and the upper surface of the horizontal part for the substrate was measured (Fig. 5f). For the angle of slope, 30, 45, and 60 degrees were chosen as experimental groups. Results of different slope angles had a similar trend for comparison of different doping concentrations and substrates, but comparison effect at 60 degrees turned to be the best and was shown here. Results for 30 and 45 degrees were listed in the supplementary information (Fig. S7, S9). It was found that the locomotion could well correspond to the material synthesis and mechanical analysis: the greater the adhesion between robot body and substrate material, the better the rolling locomotion state (faster speed and more stable posture). Moreover, microrobot rolling locomotion could overcome steep slope better (higher position and steeper angle) via porous silica doping.

Third, set up resistance with a slope for magnetically controlled trolley. If the robot was connected end to end to form a circular shape, then it could replace tires, assembling into a magnetically controlled trolley. Compared with a single robot, a magnetically controlled trolley could realize more powerful functions. Under magnetic control, it could transport and unload cargoes (Video S3). Moreover, its load ratio could exceed 40, far larger than that of common transportations (Fig. 5h), which showed the great strength of magnetic control. The slope experiments for magnetically controlled trolley also took the material type of substrate and doping concentration of porous silica as two variables, and vertical climbing height for slope angle of 60 degrees was measured (Fig. 5e, Fig. S8, S10). It was found that the greater the adhesion between robot body and substrate material, the better the magnetically controlled trolley could move forward (Fig. 5g).

From the three above experiments, it could be concluded that, for rolling locomotion and its derivative application (magnetically controlled trolley), adhesion had a positive influence, which was especially significant when encountering deficient frictional force or obvious resistance. Moreover, the porous silica doping helped with it by improving the adhesion of robot body to substrate material.

Particularly, as the essence of adhesion was attraction, when the adhesion force was too large, it was negative to mutual displacement between the microrobot and the substrate. Moreover, with the porous silica doping reducing the rigidity of microrobot, the contact surface between microrobot and substrate enlarged, which further increased the adhesion force, thus hindering the rolling locomotion. This was verified by the experiment in which dry weighing paper instead of oiled paper was taken as substrate. The dry weighing paper was rough enough and had appreciable adhesion with soft robots. Driven by magnetic force, robots with higher concentrations of doped porous silica rolled more slowly and even hard to move forward on dry weighing paper (Video S3), which gave an opposite trend compared to the results for oiled paper.

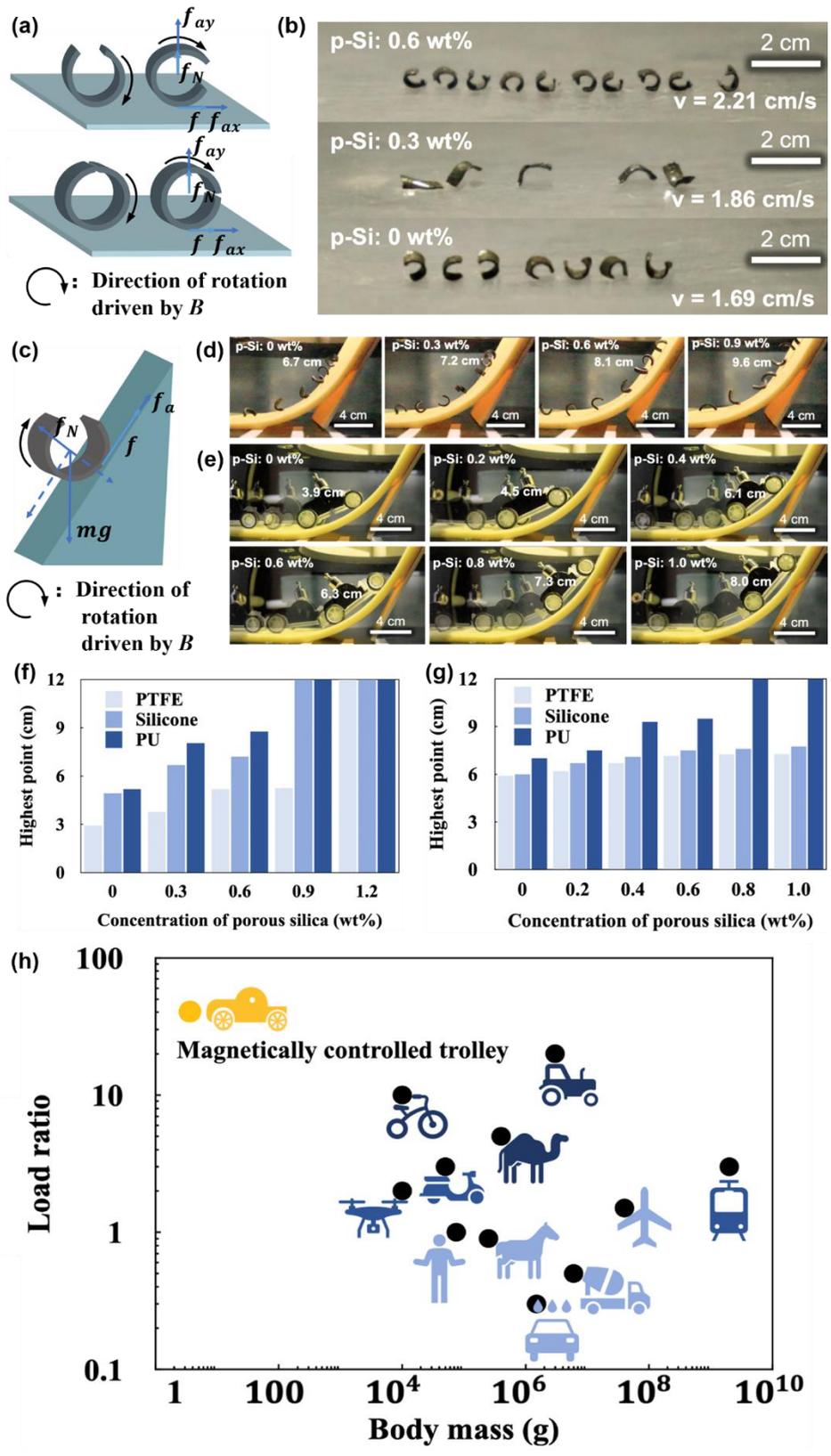

**Fig. 5.** The influence of the porous silica doping on rolling locomotion. (a) Mechanical analysis of rolling locomotion on flat surface. (b) Comparison of rolling effect on flat surface for different concentrations of porous silica doping. (c) Mechanical analysis of rolling locomotion on a slope. (d) Comparison of rolling effect on a slope for different concentrations of porous silica doping. (e) Comparison of rolling effect of magnetically controlled trolley on a slope for different concentrations of porous silica doping. (f) A bar graph showing the highest reaching point of rolling locomotion for different doping concentrations and different substrate materials. (g) A bar graph showing the highest reaching point of magnetically controlled trolley for different doping concentrations and different substrate materials. (h) An illustration of the comparison for load ratio between magnetically controlled trolley and common transportations.

### 3.2.3 Walking Locomotion, Drifting Locomotion and Magnetically Controlled Boat

Rolling locomotion was important, for it was the most effective way of marching. However, considering the small dimension of microenvironment, the speed of rolling was relatively fast. Therefore, apart from rolling, a slower way of going forward was also needed to enhance the overall control accuracy [24]. In fact, driven by a periodical magnetic field, the microrobot could walk rightward, as shown in Fig. 6b, which was the walking locomotion. This locomotion was achieved through a three-step cycle (Fig. 6a, Video S4) [1]. First, the direction of ***B*** was fixed at α = 150°, while ***B*** increased linearly until it peaked at $B_{max}$=10 mT. Concurrently, the robot was first anchored on its front end by frictional force and the horizontal component of adhesive force, and started to tilt forward. Second, the direction of ***B*** changed linearly from α = 150° to α = 120°, while maintaining ***B*** at 10 mT. As a result, the robot posed a 'C' shape. Finally, the direction of ***B*** was fixed at α = 120° while its magnitude decreased linearly from 10 mT to 0 mT. Concomitantly, the robot extended its front end, allowing it to make a net stride forward. Meanwhile, it was anchored on its back end, also by frictional force as well as the horizontal component of adhesive force. Such three steps above made up a whole walking cycle (Fig. S11).

Porous silica doping could effectively improve the walking locomotion under the circumstances of relatively small friction or large resistance, which was possibly attributed to enhanced adhesion. During the whole cycle, frictional and adhesive forces were indispensable. Without them, the robot would be bounced backward, resulting in its inability to march. Similar to previous analysis, adhesion would help with the walking locomotion, especially under the circumstances of a relatively small frictional force or large resistance. In fact, there should be a lot of fluid rushing in the human environment, which would cause great resistance to the walking locomotion of microrobot. According to the analysis above, adhesion was the key to resist adverse current and maintain walking locomotion. Under certain variables, the upper limit of adhesion determined the maximum resistance that the robot could resist. Moreover, porous silica doping could effectively increase the adhesion of microrobots to certain substrates, thus improving the stability of walking locomotion. Three groups of experiments were conducted to verify the above theory.

First, oleic acid was added to the surface of weighing paper to construct a substrate with relatively little friction for walking locomotion. Robots walked at different speeds with different concentrations if doped porous silica (Fig. 6b). Those with higher concentration had a larger adhesive force, which helped the robot to anchor its ends alternately, thus making a bigger stride in each cycle. Especially, walking locomotion was more sensitive to the lack of friction than rolling locomotion, as reflected in the more significant difference of walking speed among robots with various concentrations of porous silica doping. Moreover, the robot without doped porous silica almost marched on the spot (Fig. 6b). This was because, during walking locomotion, the contact area between the robot and the substrate was limited, so the adhesive force was deficient. This illustrated two points: adhesion was more decisive than friction; anti-resistance walking locomotion required material improvement on adhesion.

The following two experiments focused on the increase in resistance. The second experiment was setting up resistance with a slope. A slope was built to change the effect of gravity on robots' walking locomotion. Referring to the previous force analysis of rolling locomotion (Fig. 5c), increased resistance mainly consisted of the component of gravity parallel to the moving direction. In the experiment, three different materials were used as the substrate of the slope. Material type of substrate and concentration of porous silica doping were treated as two variables, and the process of robot walking on a slope was divided into three categories: reaching the top, almost reaching the top, and slipping half the way. A color block diagram was drawn to show the effect of walking locomotion in different situations (Fig. 5c). Since walking locomotion was more sensitive to resistance than rolling, the angle of 10 degrees was chosen as slope angle instead of other higher degrees, and apparent differences among testing samples could be obtained. The results correspond well to expectation: the greater the adhesion between microrobot and substrate material, the better the walking locomotion stability. Also, it could effectively improve robots' walking locomotion on a slope by porous silica doping.

The third experiment was setting up resistance with wind. A tiny fan was set to blow in the forward direction of walking locomotion, increasing its wind force at a constant speed with the time gradient (three gears in total). A piece of cloth was strung to reflect the wind force. The airflow caused great disturbance to microrobot walking locomotion, which simulated some sudden changes of air pressure and hydraulic pressure in the human environment. Similarly, the experiment took the material type of substrate and porous silica concentration as two variables, and divided the process of robot walking in the wind into three categories: successfully (reaching the end without pause and shaking), strugglingly (slight pausing or shaking, but still reaching the end), and unsuccessfully (pausing or even being blown away). A lattice diagram was drawn to show the stability of walking locomotion when encountering wind blowing (Fig. 6d), taking the concentration of doped porous silica as the abscissa and the material type of substrate as the ordinate. The results were in line with expectation: the greater the adhesion between the robot and the substrate material, the better the walking locomotion. It could be explained that adequate adhesion force kept the end of the robot well fixed on the substrate, even facing large disturbance (wind blowing).

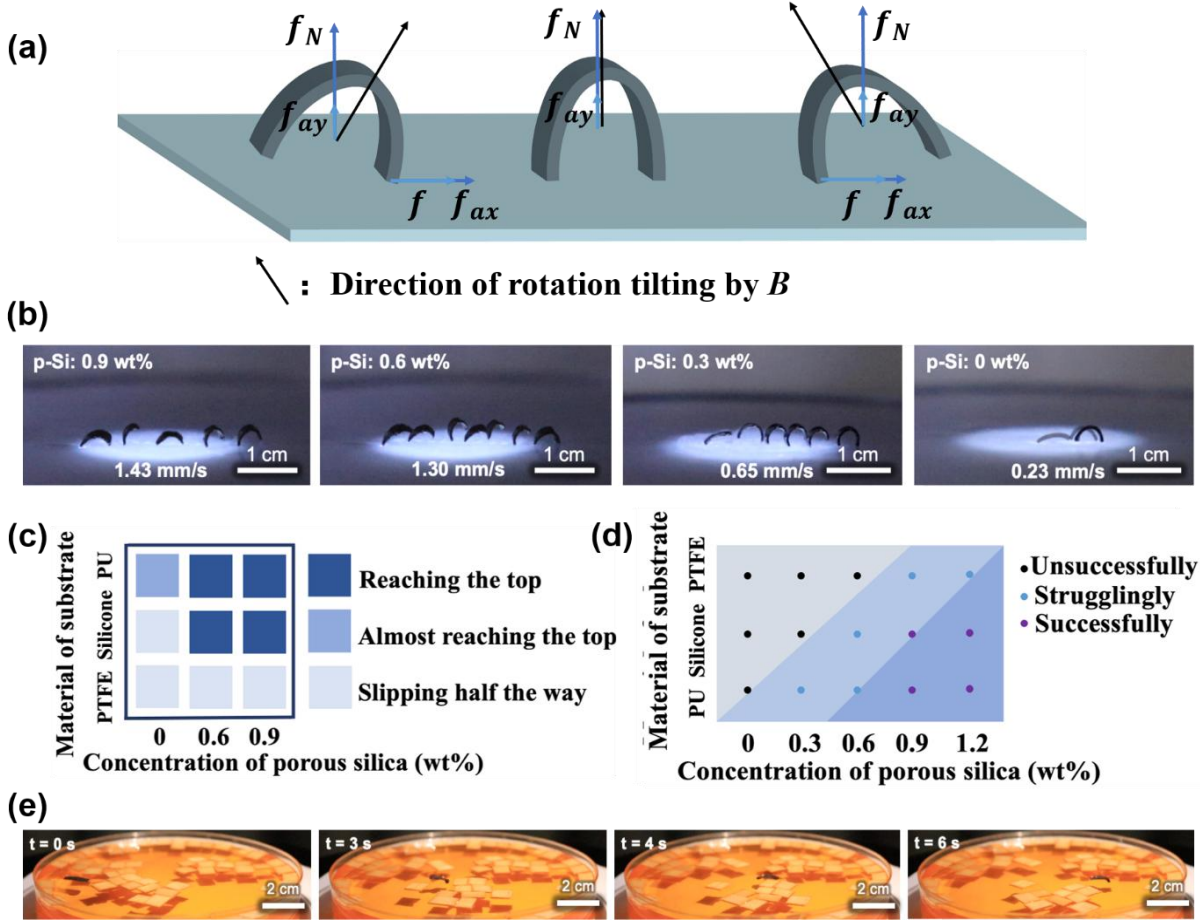

**Fig. 6.** The influence of porous silica doping on walking locomotion. (a) Mechanical analysis of walking locomotion on flat surface. (b) Comparison of walking effect on flat surface for different concentrations of porous silica doping. (c) A color block diagram showing the situation of walking locomotion for different doping concentrations and different substrate materials. (d) A lattice diagram showing the stability of walking locomotion for different doping concentrations and different substrate materials when encountering wind blowing. (e) Captures showing the process of breaking through obstacles ahead via drifting locomotion.

Based on the three experiments above, it could be concluded that porous silica doping promoted walking locomotion by improving the adhesion of robot body to substrate material, particularly when encountering deficient frictional force or obvious resistance. Since walking locomotion was another important way of marching, such results further proved that porous silica doping was greatly beneficial for the stability-enhanced marching of microrobots, because increased adhesion caused by p-Si doping was needed for both rolling and walking locomotions.

Additionally, the magnetic field setting of walking could realize the robot's drifting locomotion on water, based on reaction force provided by water and the fact that the surface tension of water was enough to allow microrobot to float. The reaction of water provided the driving force, and the forward speed was obviously larger than that of walking. There were many gas-liquid and liquid-liquid two-phase surfaces in the human environment, where the drifting locomotion was useful. Also, the robot could achieve pulling and pushing actions on the water surface, breaking through obstacles ahead (Fig. 6e). Moreover, microrobots could participate in the assembly of a magnetically controlled boat which realized more complex functions, especially cargo delivery of larger volume (Video S5).

**3.2.4 Jumping Locomotion: Vertical and Horizontal Jumping**

Jumping locomotion was important for microrobots because during the moving process, apart from contacting with substrate all the time, locomotions which could enable microrobot to move from one place directly to another place were also necessary [25]. In previous studies, leaps of robots often depended on complex mechanical drives [26]. For the magnetic soft microrobot, without complex mechanical devices, two kinds of jumping could be realized by different phase shift $\beta$, including vertical jumping (Fig. 7c, Video S6) and horizontal jumping (Fig. 7d, Video S6). However, the jumping process for magnetic soft robot was still unclear, and jumping regulations were insufficient. Here, a physical model based on quasi-static analysis was raised to analyze the essence of jumping locomotion, and the effect of porous silica doping on jumping locomotion was explained in depth.

*3.2.4.1 Vertical Jumping*

According to mechanical analysis, the relationship between rotation angle ($\theta$) and deformation along robot body ($ds$) could be expressed as below

$$B_y \cos(\omega s + \beta + \theta) - B_x \sin(\omega s + \beta + \theta) = -\frac{EI_z}{Am}\frac{\partial^2 \theta_{(s)}}{\partial s^2} \quad (1)$$

Along with boundary conditions $M_{edge} = 0$

$$\frac{\partial \theta_{(s)}}{\partial s}\bigg|_{s=0} = \frac{\partial \theta_{(s)}}{\partial s}\bigg|_{s=L} = 0 \quad (2)$$

When $\beta = -90°$, $B_x = 0$ and $B_y \neq 0$, the magnetic microrobot would perform vertical jumping locomotion. Different magnitude of $B_y$ was tried to get the corresponding deflection curves, from which the information about the mechanism of vertical jumping could be obtained (Fig. 7a).

Since $\boldsymbol{M} = \boldsymbol{m} \times \boldsymbol{B}$, initially, the left part of the robot body had the tendency of counterclockwise rotation, while the right part had the tendency of clockwise rotation. This made the middle part go up, while two ends kept in touch with the substrate, forming a small deflection convex curve. If the small deflection convex curve accorded with the final stable shape, then the robot body would keep this shape on substrate. However, if the deflection curve did not fit with the final stable shape, the robot body would undergo other kinds of deformation. When the deformation must occur but could not be accomplished relying on the support of substrate, the robot would jump up.

According to calculation, when $B_y$ was from 1 mT to 5 mT, the final stable shape was a small deflection convex curve (Fig. 7a). Therefore, the robot body would just keep this stable shape on substrate, never jumping up. In contrast, when $B_y = 6$ mT, the performance of solutions varied greatly, with the shape of deflection curve changing from convex to concave. Therefore, the small deflection convex curve formed in initial deformation was not stable, for the final state should be a concave curve with middle deflection. So the robot body had a tendency to convert to the expected final state. However, with the middle part of robot body going up, the convex curved body could not turn to concave directly with the support of flat substrate. Thus, it would continue to deform, accumulating elastic strain energy and kinetic energy while trying to reach the stable point but in fact went farther, and finally jumped up at a certain point so as to release kinetic energy and turn it to gravitational potential energy, achieving the high jump.

Nevertheless, the robot did not actually jump up when $B_y$ = 6 mT, due to the fact that $B_y$ = 6 mT was just near the critical point where the small deflection convex curve formed in initial deformation turned to be unstable, so the driving force for jumping seemed too small. Since there were other impediments for the microrobot to jump up such as frictional energy loss, such small impetus would not induce the locomotion of vertical jumping. Instead, the initial $B_y$ = 6 mT just counterbalanced the frictional energy loss (contact friction with the substrate, air drag, and viscoelastic losses within the soft robot), making the locomotion of jumping possible in stronger magnetic field.

When $B_y$ was from 6 mT to 9 mT, the degree of bending for stable deflection curve enhanced (Fig. 7a), which did not change the tendency for the robot to jump up while enhancing the impetus. When $B_y$ was larger than 9 mT, the stable deflection curve did not change apparently. However, larger magnitude of magnetic field still provided more impetus for the locomotion of vertical jumping.

In the experiment, the lowest $B_y$ for the soft robot body to jump up was 20.1 mT, which roughly equaled to the result of 18.9 mT in previous research. The magnitude of 20.1 mT accorded with our theoretical analysis above that 6 mT of magnetic field counterbalanced the frictional energy loss and provided the basic tendency for the robot to jump up, while higher magnitude of magnetic field provided more impetus for the locomotion of jumping up. In our experiment, approximately 14.1 mT of magnetic field was used to provide additional impetus (could be seen as kinetic energy) for jumping up.

According to the analysis for jumping mechanism, straight jumping height could be estimated from the perspective of energy conservation. The initial state was defined as the state where the robot body lay flat and still on substrate, and the jumping critical state was defined as the state where the robot would jump up with a little bit more magnitude of magnetic field. Based on the principle of energy conservation, between the two states, the soft robot gained energy from the work done by magnetic torque $W$. This energy was then re-distributed into three components: the change in strain energy $\Delta S$, kinetic energy $\Delta K$, and frictional losses $Q$, during the jumping process [1]. Namely,

$$W - \Delta S - Q = \Delta K \tag{3}$$

For the experiment, 6 mT of magnetic field provided the basic tendency for the robot to jump up, compensating for the energy loss, while approximately 14.1 mT of magnetic field were used to provide additional impetus for jumping up. The process could be simplified by making an assumption that the magnetic torque $W$ caused by the initial 6 mT of magnetic field all turned into frictional losses $Q$, which became the basic tendency for the robot to jump up. Then, the other part of magnetic torque $W$ caused by the sequential 14.1 mT of magnetic field turned into three parts—strain energy $\Delta S$, kinetic energy $\Delta K$, and frictional losses $Q$. The frictional losses $Q$ equaled to half of magnetic torque $W$ in this part, just as an object accelerated by a conveyor belt, only half of energy made the object to accelerate while another half became frictional losses.

According to material mechanics, the strain energy $\Delta S$ could be expressed as

$$\Delta S = \int_0^L dM = \frac{1}{2}\int_0^L EI\left(\frac{d\theta_f}{ds}\right)^2 ds \tag{4}$$

And the magnetic work done $W$ could be expressed as

$$W = -mBA\int_0^L \cos(\omega s + \theta_f)ds \tag{5}$$

At the maximum height, it was assumed that all the kinetic energy $\Delta K$ was fully converted into gravitational potential energy. In fact, part of strain energy $\Delta S$ should also turn into gravitational potential energy, but considering the shape of robot body did not change a lot, the reduction of strain energy $\Delta S$ could be omitted. Therefore,

$$\Delta K = m_r g H_{max} \tag{6}$$

Where $m_r$ represented the mass of the robot.

From the equations above, $H_{max}$ could be obtained as

$$H_{max} = 0.3507 \frac{mBA}{m_r g} \int_0^L \cos(\omega s + \theta_f) ds + \frac{EI}{2m_r g} \int_0^L \left(\frac{d\theta_f}{ds}\right)^2 ds \tag{7}$$

Assume that the stable state of the 5 mT situation was roughly the shape of robot body at the critical jumping point, after calculation (Text S3 for more detailed analysis),

$$H_{max} = 12.3 \text{ mm} \tag{8}$$

This result seemed to be a smaller approximation compared with the $H_{max}$ given by previous research, which gave the theoretical calculation value of 25.9 mm. However, it should be noticed that the actual jumping height was always lower than theoretical results, with 6.8 mm in our experiment and 4.9 mm in previous studies [1]. Therefore, in this study, the theoretical result was much closer to the real jumping height in the experiment, giving a more accurate approximation. This proved the assumption that the stable state of the 5 mT situation was roughly the shape of robot body at the critical jumping point was reasonable. Although the relative error was still non-negligible, due to the fact that the measurement and calculation were complicated for soft robot, a lot of approximation should be introduced. Previous research only showed higher relative error, but our research had made significant improvements.

*3.2.4.2 Horizontal Jumping*

In the process of vertical jumping, the forces on left and right side of the robot body were symmetric. However, in order to achieve horizontal jumping, another appropriate $\beta$ should be set so that one side could move first. According to rational design, horizontal jumping could be achieved when $\beta = 0°$ and $B_x = 0$.

The basic equation could be expressed as

$$B_y \cos(1698.16s + \theta) = -3.91 \times 10^{-9} \frac{\partial^2 \theta_{(s)}}{\partial s^2} \tag{9}$$

Similarly, different $B_y$ was tried, and the corresponding deflection curves were achieved, from which information about the mechanism of horizontal jumping could be obtained (Fig. 7b).

When magnetic field with $y$ direction was provided, the left part and right part of soft robot body had the tendency of counterclockwise rotation. In the same way, if the small deflection convex curve conformed to the final stable shape, the robot body would hold that shape on substrate. However, if the small deflection convex curve did not conform to the final stable shape, other types of deformation might occur.

According to the results, when $B_y$<32 mT, the final stable shape was only a smooth curve with small deflection, which was convex and had counterclockwise movement trend for the left part (Fig. 7b). Therefore, the robot body tended to stay in a stable state and did not jump up, achieving accumulation of elastic strain energy and kinetic energy during the process. However, when $B_y$=32mT, the stable state showed a sudden change (Fig. 7b). The shape of deflection curve on the left side changed from convex to concave, and the movement trend changed from counterclockwise to clockwise. However, at this point, direct shape change could not occur. Thus, the deformation on the left would continue until it finally left the substrate at some point and flipped to the right, where kinetic energy and elastic strain energy were converted into gravitational potential energy, achieving horizontal jump.

In actual process, $B_y$=32 mT was not the critical state for real jump due to the influence of frictional resistance. When $B_y$>32mT, the increase of magnetic field strength enhanced the accumulated energy, thus providing more power for jumping locomotion.

*3.2.4.3 Effect of Porous Silica Doping Concentration on Jumping Locomotion*

Porous silica doping had significant effects on jumping locomotion. Since experimental uncertainty was obvious for vertical jumping locomotion, it was more meaningful to analyze the relationship between porous silica doping concentration and horizontal jumping distance instead of vertical jumping height.

By comparing the jumping distance of microrobots with different porous silica doping concentrations, it could be found that, with the increase of porous silica doping concentrations (from 0 wt% to 0.9 wt%), the horizontal jumping distance increased (Fig. 7e). Namely, the horizontal jumping performance was improved by porous silica.

As was mentioned, the doping of porous silica had an effect on the mechanical properties of robot body. With the increase of porous silica doping concentration (from 0 wt% to 0.9 wt%), the elastic modulus decreased. According to the analysis above, the jumping locomotion was a turning process from elastic strain energy to gravitational potential energy. Under the same magnetic field strength, robots with different porous silica doping concentration accumulated different elastic strain energy under the same bending degree. When the concentration was 0.9 wt%, the elastic modulus was the smallest, and the jumping distance was the longest. Therefore, the essence of the effect of porous silica doping concentration on jumping locomotion was that, the change of elastic modulus caused by variable porous silica doping concentration could affect the horizontal jumping distance.

Then, the impact of elastic modulus $E$ on jumping distance was quantitatively analyzed from the perspective of energy conservation. Similar to vertical jumping, the transformation of magnetic torque $W$ to strain energy $\Delta S$, kinetic energy $\Delta K$ and friction loss $Q$ was also carried out in the process of horizontal jumping (Equation 3).

In order to simplify the process, it was assumed that friction loss during the horizontal jumping process with different porous silica doping concentrations was the same. Since the motion trajectories of different microrobots were basically the same, this hypothesis could be considered reasonable. For strain energy $\Delta S$, the expression had previously been derived as Equation 11. Therefore, as $E$ decreased, strain energy $\Delta S$ also decreased. Under the same magnetic moment, the larger the portion of energy converted to kinetic energy $\Delta K$, the greater the initial velocity of jumping, resulting in a longer jumping distance. This was consistent with experimental results.

Therefore, porous silica doping had a beneficial effect on possible practical applications of microrobots by enhancing the horizontal jumping distance. Namely, microrobots with porous silica doping could achieve farther places due to softer body and higher flexibility, which indicated they might accommodate better to a complex environment by using jumping locomotion when the surroundings were not suitable for walking or rolling.

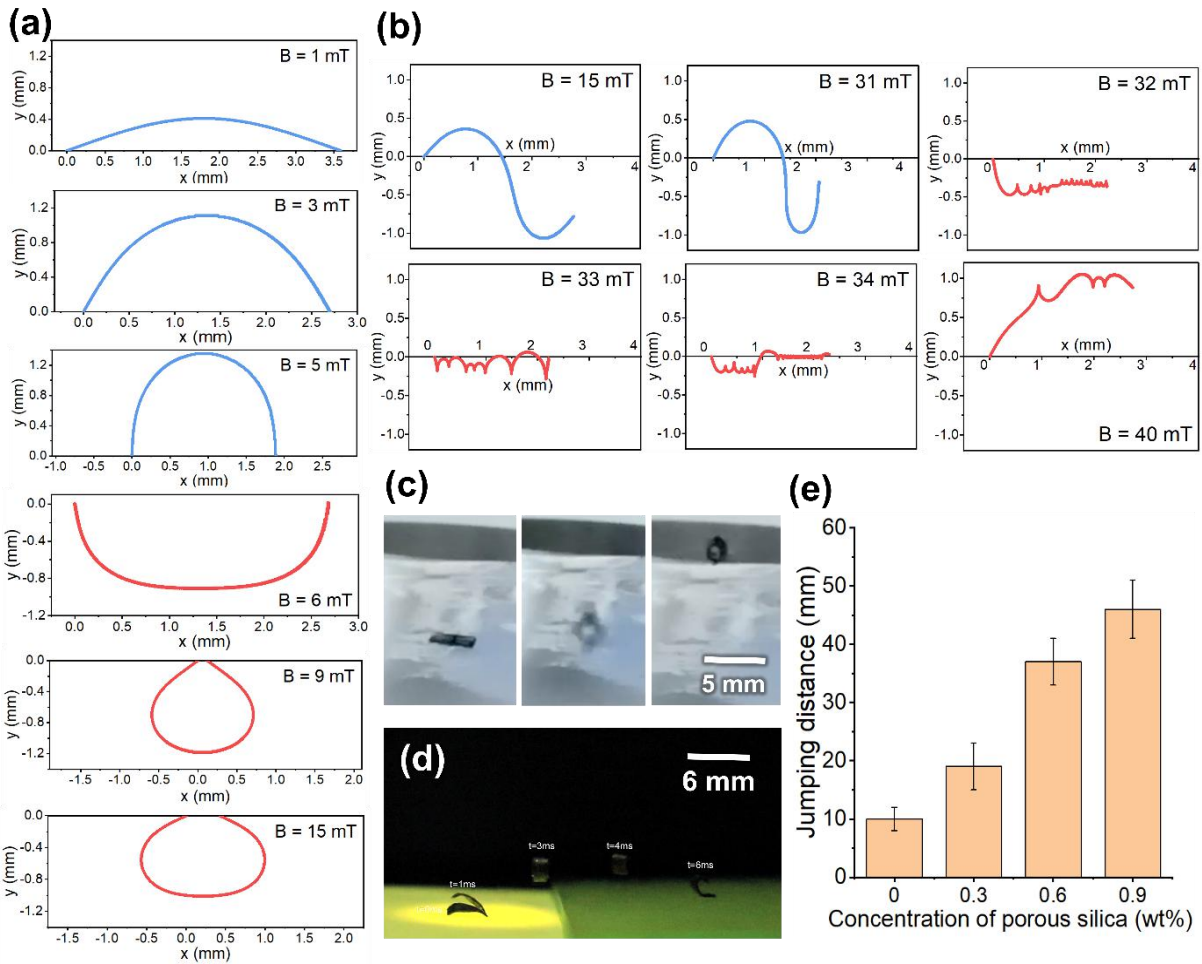

**Fig. 7.** Analysis for jumping locomotion, including vertical jumping and horizontal jumping. (a) Deflection curves of robot body in different magnitude of magnetic fields (1, 3, 5, 6, 9, and 15 mT) for vertical jumping locomotion, showing the mechanism of vertical jumping. (b) Deflection curves of robot body in different magnitude of magnetic fields (15, 31, 32, 33, 34, and 40 mT) for horizontal jumping locomotion, showing the mechanism of horizontal jumping. (c) Captures showing the process of vertical jumping

locomotion. (d) Captures showing the process of horizontal jumping locomotion. (e) A graph showing the relationship between porous silica concentration and horizontal jumping distance.

**3.2.5 Crawling Locomotion and Thrombus Removal**

For practical applications, microrobots should not only move on a surface, but also move in a confined space, such as fine pipes in chemical reactors and animal blood vessels.[27] In such environments, microrobots could not travel freely, and the corresponding locomotion must be realized by collision with the sidewall. A direct idea was that if a periodic rotating magnetic field was applied, the robot could hit the pipe wall periodically, so it was possible to move forward continuously with a cosine shape (Fig. 8a-b, Video S7). This was similar to the locomotion of undulating swimming [28]. However, swimming was just an analogy, which could not be directly used to analyze crawling locomotion due to significant differences in boundary conditions. For the crawling of magnetic soft microrobot, previous studies only gave a fitting modal instead of clarifying the physical mechanism [1]. Here, a physical model based on analogy to mechanical wave was raised to get closer to the essence of crawling locomotion, and the effect of porous silica doping on crawling locomotion was explained in depth. Moreover, demonstration of thrombus removal was conducted to indicate application prospects of crawling locomotion, and porous silica doping might provide a stronger impetus for thrombus removal.

*3.2.5.1. Analysis of Crawling Mechanism*

According to the general theory, small deformation of microrobot could be expressed as:

$$y_{(x)} = \frac{mBS}{EI_z\omega^3} cos\omega x, \quad 0 < x < L \tag{10}$$

In the crawling process, at any fixed time, the microrobot showed the shape of a cosine wave function (Fig. 8b, Video S7), just like the mathematic relationship given in Equation 9. However, considering time as another variable, the crawling locomotion did not work like either a normal traveling wave or a normal standing wave, instead, somewhere in between (Text S4 for more detailed analysis). Meanwhile, the amplitude $A = \frac{mBS}{EI_z\omega^3}$, which was only suitable in free propagation, became no longer feasible owing to the boundary of glass tunnel. Therefore, in order to describe the crawling locomotion, a new amplitude $A' = 0.6$ mm should be adopted, which was the radius of glass tunnel. Thus, the wave function describing crawling locomotion could be expressed as

$$y_{(x,t)} = A'cos(\omega x + 2\pi\gamma f t), 0 < x < L \tag{11}$$

$\gamma$ was a damping coefficient that implied how the propagation of wave was influenced by the tube wall. $f$ was the wave frequency, which equaled to the frequency of the applied rotating magnetic field. The influence of gravity was negligible in this case. To obtain the value of $\gamma$, the free amplitude A should be calculated first, by substituting $m = 77500$ A·m$^{-1}$, $B = 0.020$ T, $S = wh = 2.78 \times 10^{-7}$ m$^2$, $E = 9.30 \times 10^4$ Pa, $I_z = 7.91 \times 10^{-16}$ m$^4$, $L = 0.0037$ m. Thus,

$$A = \frac{mBS}{EI_z\omega^3} = \frac{mBSL^3}{8EI_z\pi^3} = 1.20 \times 10^{-3} \text{ m} \tag{12}$$

Then the coefficient $\gamma$ could be expressed as the cubic of the ratio of bounded amplitude and free amplitude

$$\gamma = (\frac{A'}{A})^3 = 0.125 \tag{13}$$

Since the robot body was seen as a whole period, and wavelength $\lambda$ was just microrobot body length $L$, the speed of robot $v$ equaled to the wave speed $u$. Considering the phase propagation delay coefficient $\gamma$, the actual speed of microrobot could be expressed as

$$v = \gamma L f \tag{14}$$

Substituting $\gamma = 0.50$, $L = 3.7$ mm and $f = 15$ Hz, it could be calculated that

$$v = 6.94 \text{ mm·s}^{-1} \tag{15}$$

The speed of microrobot $v$ indicated how the robot crawled in a glass tunnel with a rotating magnetic field.

*3.2.5.2. Effect of Porous Silica Doping Concentration on Crawling Locomotion*

By doping different concentrations of porous silica in robot body, the corresponding Young's modulus could be obtained by tensile test (Fig. 3h). For the doping concentration of 0 wt%, 0.3 wt%, 0.6 wt%, 0.9 wt%, and 1.2 wt%, the calculated speed of the robot body $v$ in the same rotating magnetic field was 6.94 mm·s$^{-1}$, 5.40 mm·s$^{-1}$, 2.18 mm·s$^{-1}$, 0.98 mm·s$^{-1}$ and 4.11 mm·s$^{-1}$, respectively (Fig. 8c), which roughly conformed to the experimental results (3.01 mm·s$^{-1}$, 3.20 mm·s$^{-1}$, 2.12 mm·s$^{-1}$, 1.13 mm·s$^{-1}$ and 4.30 mm·s$^{-1}$, respectively, lower than expectation at 0 wt% and 0.3 wt% due to low friction-induced slippery without enough porous silica doping). Hence, the crawling speed of the robot body had a trend of decreasing-increasing with the increasing of porous silica doping concentration, which meant that different porous silica concentrations could be chosen for different needs for crawling speed. For maximum speed, 1.2 wt% concentration turned to be the best. For minimum speed, 0.6-0.9 wt% concentration was proper. 0-0.3 wt% concentration was not suitable for crawling because of low friction-induced slippery, which further proved that porous silica doping could enhance locomotion stability by enhance adhesion to substrate.

*3.2.5.3. Application of Crawling Locomotion*

Crawling locomotion was promising for minimally invasive surgery, and the application in surgical thrombectomy was remarkable. Blood circulates normally in the bloodstream, but in some cases, blood might clump to form thrombus, as known as blood clots. Having life-threatening blood clots in the body might block the blood vessels, arteries, and veins, causing angiemphraxis, which might suddenly take away a human's life without any signals. Having a thrombectomy was the only way, but there were a few risks that the patient might face like excessive bleeding, permanent damage in blood vessels and pulmonary embolism [29]. Hence a minimally invasive way should be adopted, and microrobot with crawling locomotion was a representative example (Fig. 8b, Video S7). The images below illustrated the whole process of thrombus removal. Wet thrombus as well as dry thrombus was studied respectively. In the wet thrombus case, microrobot was put in a glass tunnel with a radius $r = 0.6$ mm. From t = 4 s to t = 9 s, the robot crawled forward and met the wet bionic thrombus. From t = 9 s to t = 627 s, the robot successfully pushed away from the thrombus and crawled over the glass tunnel. In the dry thrombus case, the robot could finish the task in 1.70 seconds, which was significantly faster than the wet case comparatively. By doping different amounts of porous silica, the corresponding forces that microrobot push the thrombus could be calculated by the momentum theorem

$$Ft = mv \qquad (16)$$

$t$ was time of interaction, m was the mass of a microrobot, and $v$ was the speed of crawling. It was noteworthy that $t$ equaled to the period $T$, because it could be assumed that the robot body would completely stop after each collision, which was reciprocal of frequency $f$. Hence,

$$F = mvf \qquad (17)$$

By substituting the mass $m = 1.91 \times 10^{-6}$ kg, and frequency $f = 15$ Hz, the corresponding forces that microrobots doped by porous silica of 0 wt%, 0.3 wt%, 0.6 wt%, 0.9 wt%, and 1.2 wt% interacted with the thrombus could be obtained as $1.99 \times 10^{-7}$ N, $1.55 \times 10^{-7}$ N, $0.625 \times 10^{-7}$ N, $0.281 \times 10^{-7}$ N, $1.18 \times 10^{-7}$ N respectively (Fig. 8d), which roughly conformed to the experimental results ($0.862 \times 10^{-7}$ N, $0.917 \times 10^{-7}$ N, $0.607 \times 10^{-7}$ N, $0.324 \times 10^{-7}$ N, and $1.23 \times 10^{-7}$ N respectively, lower than expectation at 0 wt% and 0.3 wt% due to low friction induced slippery without enough porous silica doping). Hence, the pushing force of microrobot had a trend of decreasing-increasing with the increasing of porous silica doping concentration, which meant that different porous silica concentrations could be chosen for different needs for thrombus removal. For maximum force, 1.2 wt% concentration turned to be the best. For minimum force, 0.6-0.9 wt% concentration was proper. 0-0.3 wt% concentration was not suitable for crawling because of low friction-induced slippery, which further proved that porous silica doping could enhance locomotion stability by enhancing adhesion to the substrate. In most cases, a force large enough was needed for thrombus removal, and then 1.2 wt% concentration would be a proper choice.

In summary, the crawling locomotion of magnetic soft microrobot was clarified in this study. The essence of the crwaling could be seen as a mechanical wave with phase propagation delay due to pipe wall constraints. Flexibility was very important for motion under boundary constraints, which highlighted the significance of porous silica doping. Microrobots doped with 1.2 wt% porous silica could move the most smoothly in a tube and had the potential for thrombus removal due to relatively strong impact force.

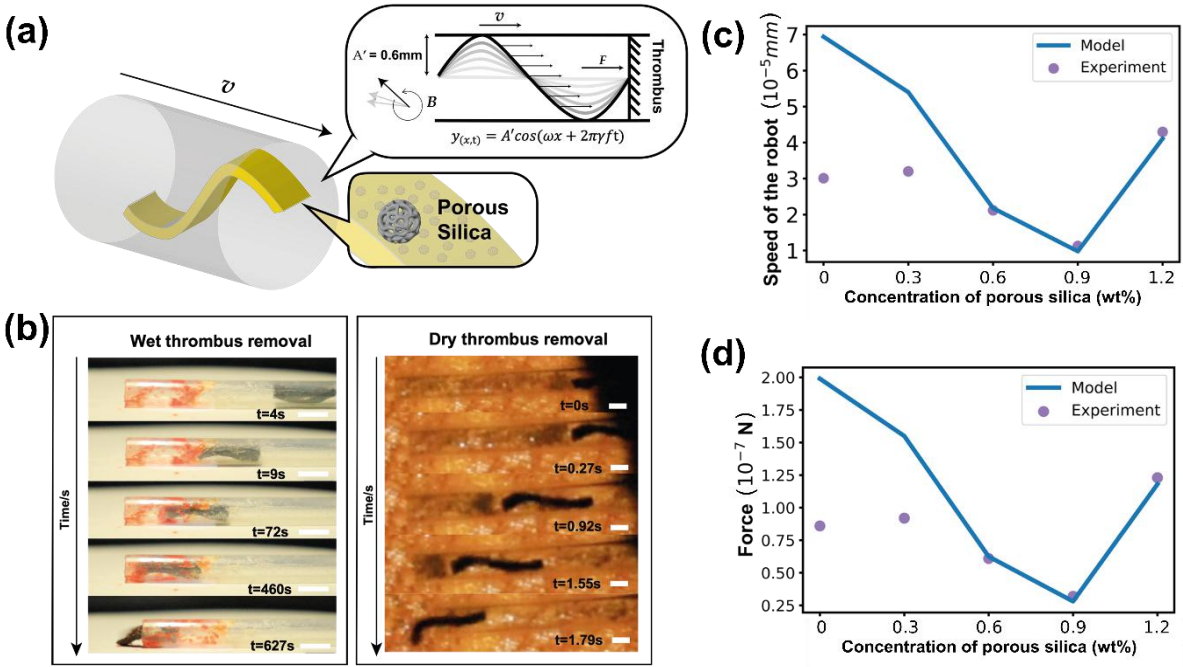

**Fig. 8.** Schematic of crawling locomotion and thrombus removal. (a) An illustration of the crawling mechanism. (b) Demonstration of wet thrombus removal as well as Dry thrombus removal. Scale bars, 5 mm for the left and 1 mm for the right. (c) A graph illustrating the relationship between crawling speed (pushing force) and porous silica doping concentration. (d) A graph illustrating the relationship between pushing force and porous silica doping concentration. In (c) and (d), theoretical results and experimental results were shown simultaneously.

### 3.2.6 Stirring Locomotion and Biocatalysis

*3.2.6.1 Stirring Mechanism*

By applying a rotating magnetic field with three-axes-alternating (Fig. 9a), the robot stirred the solution steadily (Video S8). The mechanism of stirring locomotion was similar to rolling locomotion, in other words, rolling in liquid environment. Stirring locomotion could highly increase the rate of reaction as it could make reactants mixed more intensively [30]. Compared with a normal magnetic stirrer, microrobot could be made into any size conveniently, and its soft body made it perform adaptive locomotion [31], not to be stopped by hitting the container wall, but showed a stronger stirring effect by circulating along the container wall instead. From this perspective, the porous silica doping concentrations which could soften the robot body would contribute to stirring locomotion, enhancing locomotion stability by promoting its adaptive locomotion.

Owing to the features of the robot-like small dimension and magnetic control, it was promising for organizing an intelligent microreactor (Fig. 9b). The model was assembled by a wooden platform and plastic container. The robot could walk over the platform and jump into the container, which played a role of a reactor. Then the robot started stirring and rolled away after finishing the task (Video S8). By using this combination of walking, stirring, and rolling locomotion, an efficient intelligent microreactor was achieved. Such an intelligent microreactor could proceed with a reaction automatically without touching the reactants inside, contributing to the remote control of chemical reactions and optimization of reaction conditions [32,33].

*3.2.6.2 Applications on Intelligent Microreactor and Biocatalysis*

Meanwhile, there was also a promising application in biocatalysis by microrobots loading various enzymes. Enzyme loading was an important direction for biocatalysis because it could greatly facilitate targeted catalysis and enzyme recycling [34,35]. Enzyme-loading microrobots could work as catalysts and magnetic stirrers at the same time, enhancing the reaction rate by two mechanisms together. The glucose oxidase (GOx) and horseradish peroxidase (HRP) cascade reaction was taken as an example, which had prospects for cancer starvation therapy as well as high-throughput detection of blood glucose [36,37]. Glucose was oxidized to gluconic acid with the catalysis of GOx, and forming $H_2O_2$, which would be decomposed into two hydroxyl radicals. Eventually, hydroxyl radicals reacted with ABTS, forming $ABTS^{-\cdot}$ whose solutions presented blue color and characteristic absorbance was between 400 nm to 430 nm [38]. Two different enzymes catalyzed two consecutive steps of reactions, with blue color occurring in the final step (Fig. 9c). In this research, GOx and HRP were loaded in the base material of microrobots. After magnetization, the enzyme-loaded robot was put into a solution with glucose and ABTS in a rotating magnetic field. After a few seconds, the solution significantly turned blue, verifying the catalysis effect of enzyme-loading microrobots.

Moreover, the enzyme-loading microrobots had good performance in renewability after using an abrasive paper to rub the robot surface (Fig. 9d). From the first usage to the fourth usage, the average reaction rate from t = 0 s to t = 240 s was $1.19 \times 10^{-4}$, $1.03 \times 10^{-4}$, $6.70 \times 10^{-5}$ and $2.19 \times 10^{-5}$ mmol·$L^{-1}$·$s^{-1}$ respectively, which revealed that the enzyme-carrying robots still had significant biocatalysis properties after repeated usage.

A stirring microrobot was also beneficial in heterogeneous reactions (Fig. 9e). At t = 0 s, 20 µL of mixed enzyme aqueous solution containing GOx as well as HRP was added, which floated on the surface of glucose-ABTS-glycol (50%) mixed solution, forming a phase interface with blue color [39]. The moving of phase interface was extremely slow with significant moving after 53 seconds, which greatly limited the reaction rate. However, when stirring locomotion was triggered, reactants were immediately mixed fully so the reaction could proceed smoothly.

Controlling the process of reaction was also an essential part, because it was necessary to make reactions proceed stepwise and controllable. The GOx-HRP cascade reaction could be set as a representative example via controlling stepwise. By loading GOx and HRP in two microrobots, respectively, the stepwise reaction could be achieved automatically (Fig. 9e,

Video S8). First, the robot loading GOx rolled into the reactor. After 79 seconds of stirring, the first robot rolled out, and the first step of GOx-HRP cascade reaction finished, forming $H_2O_2$. Then the second robot loading HRP rolled in and began to stir, and then the solution started to turn blue gradually, which was the second step of GOx-HRP cascade reaction.

In summary, the stirring locomotion of magnetic soft microrobots, derived from rolling locomotion, could work as flexible magnetic stirrers. Since porous silica doping could enhance flexibility, microrobots doped with porous silica stirred in a confined container more smoothly. Apart from stirring, microrobots could also carry enzymes for targeted biocatalysis. Combined with multimodal locomotion, such microrobot could contribute to the core of an intelligent microreactor, realizing remote and automatic control of chemical reactions.

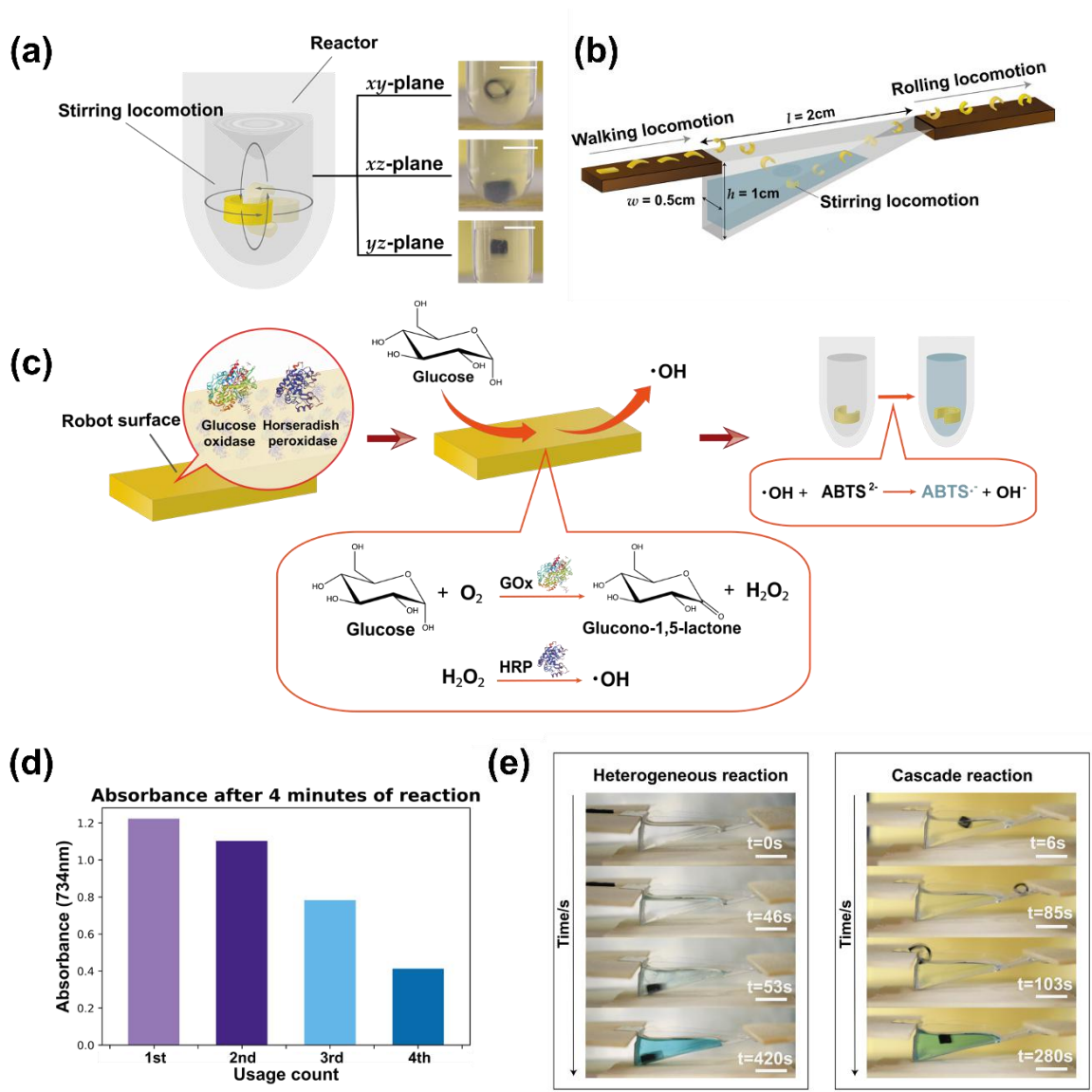

**Fig. 9.** Schematic and data graph of stirring locomotion and biocatalysis. (a) An illustration of the stirring locomotion. Scale bars, 0.5 cm. (b) An illustration of the intelligent microchemical engineering system. (c) An illustration of the enzyme carrying microrobot for biocatalysis, namely, the GOx-HRP cascade reaction. (d) A graph showing the relationship between usage times and absorbance after 4 minutes of reaction, showing the renewability of the enzyme-carrying microrobot. (e) Captures showing the process of heterogeneous biocatalysis reaction and cascade biocatalysis reaction controlled by stirring locomotion of microrobots in the intelligent microchemical engineering system. Scale bars, 0.8 cm.

### 3.2.7 Simulation of Multimodal Locomotion *in Vivo*

To illustrate the robot's ability to actualize multimodal locomotions and tasks assigned in real situation, microrobots were put under simulated *in vivo* environment (Video S9). In order to fully explore the robot's potential at realistic significance, representative organ models in surgical process were selected as the background environment: human stomach (Fig. 10a, 10b) and blood vessel (Fig. 10c, 10d). In accordance to the realistic anatomical situation where body fluid was randomly distributed in an organ, both conditions as of with and without fluid were simulated.

The robots performed different combinations of motions in response to different environments. On rugged paths such as in the stomach, a complex combination of modes of motions was applied in order to overcome the obstacles. In addition, compared to the watery environment, the waterless path had more resistance to movement. A varied multimodal series of motion (including walking, jumping, and rolling), therefore, was required to go over the dynamic barrier of the randomly protruding surfaces (Fig. 10a). During the walking locomotion, the robot tilted its gravity center and extended its front leg across the low-height wrinkled surface to go forth. By shifting anchorage back and forth, it came across the mildly rugged surface. When the robot met a barrier of greater height, the magnetic field was adjusted to induce shape deformation and rigid body rotation, so that the robot jumped over the high barrier. On relatively smooth terrain, ***B*** rotated clockwise to produce rolling locomotion, increasing the efficiency of movement.

On the watery surface where fewer obstacles on the forwarding path were present, the robot moved with more suitable multimodal locomotion: drifting and rolling (Fig. 10b). The surface tension of liquid supported the robot's movement, allowing it to stand and walk on the water surface. With basic locomotion similar to walking but on the fluid surface, the robot drifted off at relatively high speed. The robot could also roll in the watery environment, which was at the high efficiency of delivery as well. ***B*** changed into a rotating magnetic field, bending the robot to a circular curved shape that allowed for rolling. As it started with the rolling locomotion, which generated greater disturbance to the environment, surface tension was no longer capable of supporting its motion on surface. The robot, thus, started rolling at the solid base of the environment.

As the environment was relatively uniform in the simulated blood vessel, a simpler combination of motion was applied. With a steady surface and few obstacles on the path, a rotating *B* was applied to induce rolling locomotion for higher efficiency of delivery (Fig. 10d). By the adjustment of magnetic field rotating plane, microrobot could turn in the rolling process, so as to adapt to the complex bifurcation structure in blood vessels. Moreover, two locomotions (drifting and rolling) in the fluid-contained blood vessel model were realized together (Fig. 10c), demonstrating the robot's ability to move both on and under water surface. Both of these locomotions, controlled by the changing magnetic field, were effective transferring methods of high speed in watery environments.

In order to simulate movement in the blood vessel with greater consistency to reality, water current was applied in the scenario to demonstrate body fluid with high fluidity (Video S9). The water currents were a simulation of random body fluid flow. Under rotating *B*, the robot rolled back and forth over the created path where water flowed through, illustrating its ability to move steadily, mostly unaffected by the surrounding flow field. In addition, the simulation of muscle peristalsis was also realized, and the rolling locomotion could still take place in such an environment (Video S9).

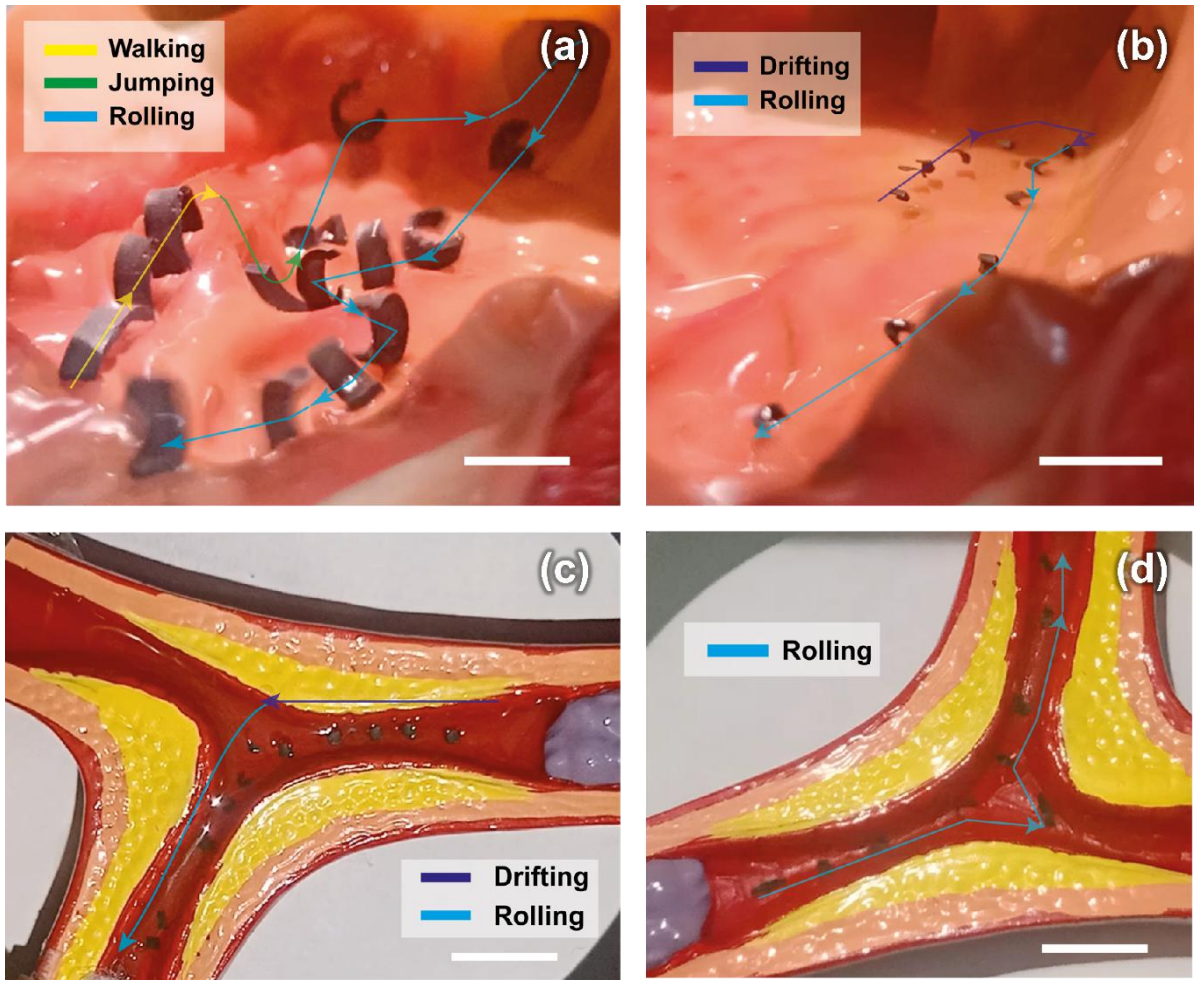

**Fig. 10.** Simulation of multimodal locomotion *in vivo*. (a) A simulation of multimodal locomotion in human stomach, solid environment. Scale bar, 8 mm. (b) A simulation of multimodal locomotion in human stomach, watery environment. Scale bar, 8 mm. (c) A simulation of multimodal locomotion in blood vessel, watery environment. Scale bar, 2 cm. (d) A simulation of multimodal locomotion in blood vessel, solid environment. Scale bar, 2 cm.

From the simulation above, it was concluded that microrobots in this study had the potential for *in vivo* movement. Porous silica doping enhanced microrobot locomotion stability by increasing adhesion as well as softening the robot body, thus enlightening the prospect of multimodal locomotion in the nonideal environment with great disturbance. Compared with previous research, this study contributed significant progress. For multi-tasking precision medicine, microrobots would certainly meet a lot of nonideal hindrances, which was not placed in a vital position for previous studies. However, no matter how sophisticated single locomotion was, if it could not proceed normally in actual working environment, then the design would be greatly limited for future use. In this study, every single locomotion was proved to have enhanced stability via porous silica doping, and they could still take place normally when combined together in a simulated *in vivo* environment with original as well as artificial hindrance. Thus, the prospect of magnetic soft microrobot for precision medicine became closer to reality.

## 4. Conclusions

In conclusion, a porous silica doped flexible magnetic microrobot for enhanced stability of multimodal locomotion in nonideal biological environment was demostrated in this study (Fig. 11). The doping of porous silica not only improved adhesion properties but also refined comprehensive mechanical properties of the microrobot. On the basis of fundamental locomotions, including rolling, walking, jumping, and crawling, and complex locomotions combined by basic motions via rational design, including drifting, stirring, trolley, and boat, the influence of porous silica doping on the stability of each locomotion in nonideal environment was explored in depth. The porous silica doping turned out to be benificial for stability enhancement of most locomotions. Diffenent from previous locomotion demonstrations in ideal environment, due to porous silica doping, motions in this study could be conducted in nonideal circumstances such as climbing, loading, current rushing, wind blowing, and obstacle hindering. Such stability-enhanced multimodal locomotion system was used in biocatalysis as well as thrombus removal, and *in vivo* demonstration of multimodal locomotion with nonideal disturbance enlightened its prospect for precision medicine.

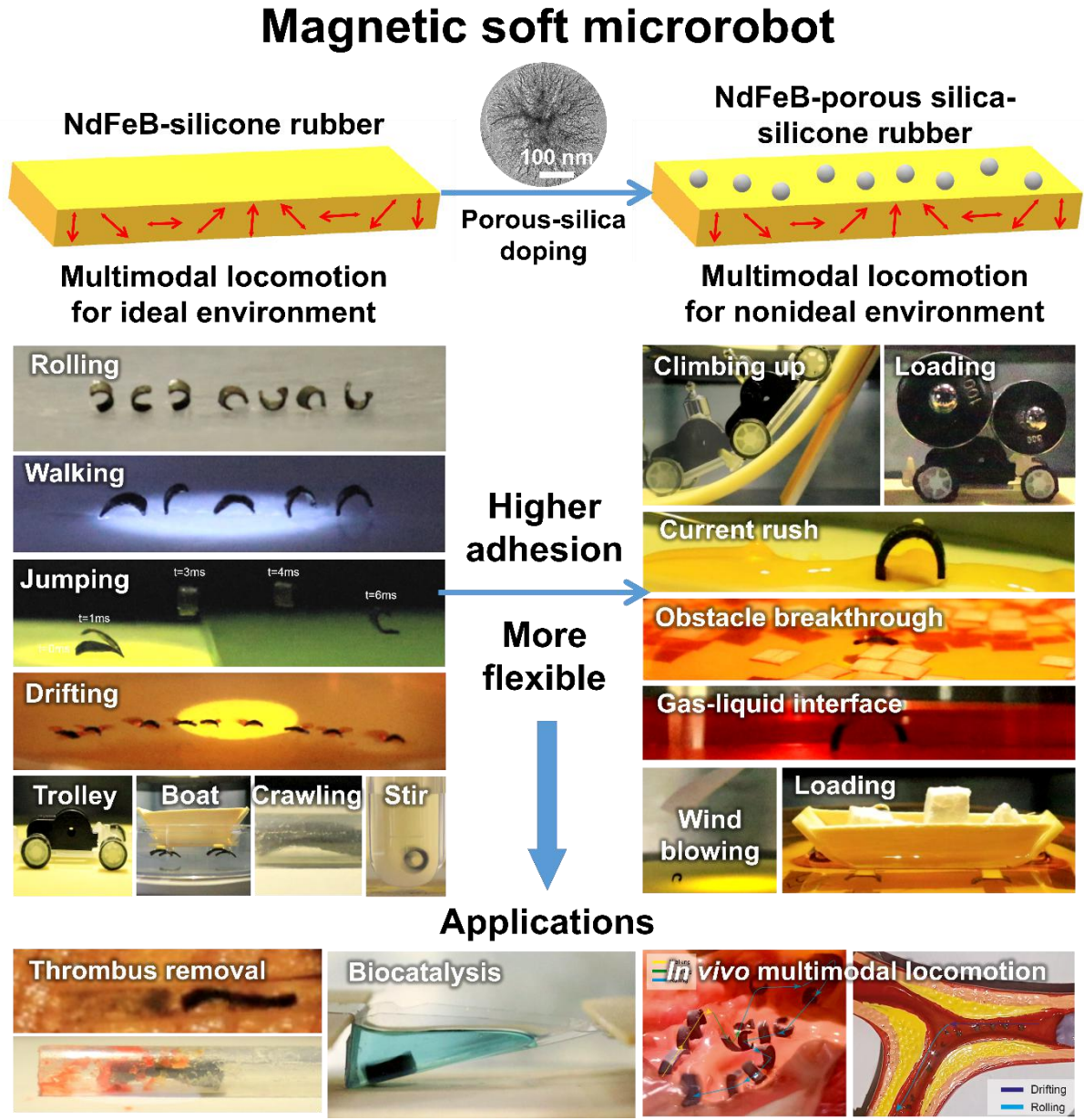

**Fig. 11.** The summary figure, an overall illustration of the porous silica doped soft magnetic microrobot for enhanced stability of multimodal locomotion in nonideal biological environment.

However, this study still has obvious limitations. Compared with previous reasearch, this work has innovatively made an attempt to explore locomotion stability in pratical nonideal biological environment. However, since the concept of nonideal biological environment itself is too complex to be explained by limited experiments, this work could only focus on some major kinds of distubance that microrobot may face in human body. Thus, relative results only indicate a theoretical feasibility instead of assuring the practicability in real medical application. Currently, the development of magnetic soft microrobot is still at the state of beginning, and *in vivo* experiment is hard to conduct (although appropriate simulation can be conducted) for sake of safety and ethics [40]. After more demonstration for security, *in vivo* experiment will be conducted to verify the feasibility of this porous silica doped flexible magnetic microrobot for enhanced stability of multimodal locomotion in nonideal biological environment, contributing to the future develoment of precision medicine and other related aspects.

## Supporting Information

Supporting Information is available online.

## Acknowledgements

This work was supported by the National Key R&D Program of China (2018YFA0901700), the National Natural Science Foundation of China (21878173), the Beijing Natural Science Foundation (2192023), and a grant from the Institute Guo Qiang, Tsinghua University (2019GQG1016).

## Author Contributions

S. Li, D. Liu and Y. Lu proposed and designed the research. S. Li, Y. Hu, Z. Su, X. Zhang, D. Li and R. Guo performed the experiments. S. Li, Y. Hu, Z. Su and D. Li developed theoretical models. D. Liu, Y. Lu, and R. Guo provided instructions on experimental design and paper writing. S. Li, Y. Hu, Z. Su, X. Zhang and D. Li analyzed experimental data and wrote the paper. X. Zhang finished video capture and editing. All authors participated in discussions.

**Conflict of Interest**

The authors declare no conflict of interest.

# Supplementary Information

**Soft magnetic microrobot doped with porous silica for stability-enhanced multimodal locomotion in nonideal environment**

*Shangsong Li [a,b,#], Dong Liu [a,#], Yvping Hu [a,#], Zhijie Su [a], Xinai Zhang [a,c], Ruirui Guo [a,d], Dan Li [a], Yuan Lu [a,*]*

[a] Key Laboratory of Industrial Biocatalysis, Ministry of Education, Department of Chemical Engineering, Tsinghua University, Beijing 100084, China.

[b] School of Materials Science and Engineering, Tsinghua University, Beijing 100084, China.

[c] Beijing No.4 High School International Campus, Beijing, 100034, China.

[d] College of Bioengineering, Tianjin University of Science and Technology, Tianjin, 300457, China.

# These authors contributed equally to this work.

* Corresponding author. E-mail: yuanlu@tsinghua.edu.cn

**Experimental Section**

*Chemicals*

Triethanolamine (TEA), hexadecyl trimethyl ammonium bromide (CTAB), sodium salicylate (NaSal), tetraethyl orthosilicate (TEOS), diammonium ABTS, glucose, glucose oxidase (GOx), horseradish peroxidase (HRP), NdFeB magnetic microparticles, oleic acid, ethylene glycol, and ethanol were purchased from Sigma-Aldrich. Ecoflex 00-10 polymer matrix was purchased from Smooth-On Inc. Silicone rubber, polyurethane (PU), and polytetrafluoroethylene (PTFE) were purchased from DuPont Chemical Co., Ltd. Deionized water was purified using a Milli-Q system. All other chemicals were used as received.

*Synthesis of Dendritic Mesoporous Silicon Nanoparticles*

60 μL TEA was added to 25 mL deionized water, and the mixture was shaken at 160 rpm in a water bath at 80 °C for 0.5 h. Then 380 mg CTAB and 168 mg NaSal were added and continued shaking was applied for 1 h. Afterward, 4 mL TEOS was added, and a continued shaking was applied for 2 h. Then the mixture was centrifuged at 7000 rpm for 20 min to collect the products. The products were washed with ethanol for three times to remove the residual reactants, and finally dried overnight in a vacuum at room temperature.

*Preparation of Magnetic Soft Microrobot*

The composition of the robot involved neodymium iron boron (NdFeB) magnetic micro-particles (MQP 15-7, Magnequench; average diameter: 5 μm, density: 7.61 g/cm3), the Ecoflex 00-10 polymer matrix A and B (Smooth-On Inc.; density:1.04 g/cm3), and porous silica (p-Si) microparticles. The amount of NdFeB, Ecoflex A, and Ecoflex B added for each glass slab was determined in correspondence to the thickness of the layer and area of the slab, and the amount of porous silica added was calculated through the demanded concentration of the material.

Two specifications were made for different uses throughout the experiment. The first specification created films of 1mm. Slabs of approximately 100 cm$^2$ were used as a model for film construction. Ecoflex A (4.4 g) was added onto the glass slab first, and then NdFeB micro-particles (8.8g) were added onto the polymer matrix Ecoflex A on slab in order to avoid dispersion to other places. Porous silica micro-particles were added in the calculation of concentration (for example, the preset 0.2% concentration involved 0.0352 g of porous silicon micro-particles). Ecoflex B (4.4 g) was added at last, since it started to react with Ecoflex A to form crosslinking structure once added. The second specification created films of 0.2 mm, which involved glass slab of 48 cm$^2$ and the addition of 1 g NdFeB micro-particles, 0.5 g Ecoflex A, 0.5 g Ecoflex B, and porous silica particles in correspondence to concentration.

After the addition of materials, the stirring glass rod was used to fully mix the mixture on the glass slab and to ensure that film of uniform thickness was made. The uniformly mixed mixture was placed in the air for 5 hours to solidify.

*Measurements and Characterization*

SEM images were recorded on a Zeiss Gemini high-resolution scanning electron microscope (Germany) operating at 5 kV. The EDS images were recorded while operating at 15 kV. Microrobot samples were first put into liquid nitrogen and then broke, leaving quenched sections for observation. TEM images were taken with a JEOL 2011 microscope (Japan) operated at 200 kV. Magnetic characterization was conducted on a vibrating sample magnetometer LakeShore 8604.

*Magnetization of Magnetic Soft Microrobot*

After solidification, the film was cut into pieces, becoming the robot bodies. The cutting involved two specifications, each corresponding to the specific thickness of the film. For films of 1 mm (h=1 mm), each robot had dimensions of w=3.6 mm and L=8.9 mm; for films of 0.2 mm (h=0.2 mm), each robot had dimensions of w=1.5 mm and L=3.7 mm.

The robot body of 8.9 mm length was then wrapped around a cylindrical wood rod with a circumference of 8.9 mm; the robot with 3.7 mm length was wrapped around a rod with 3.7 mm circumference. In order to receive uniform magnetization, the robots were vertically wrapped around the rod without any overlapping. The soft-bodied robots were then put into the pulse magnetizer (Beijing EUSCI Technology Limited), which could give a magnetic pulse of 2.5 T for magnetization. The angles by which robots were put into the magnetizer turned to be important, as it necessarily created a phase shift $\beta$ in the magnetization profile $m$. It was determined by the angle between the vertical direction and the wrapping start point. After magnetization, the robot was unwrapped, and a harmonic magnetization profile was achieved.

*Tensile Test and Tensile-Adhesion Test*

Tensile test of materials containing different concentrations of porous silica was performed to directly reveal the mechanical properties of the synthesized material. The tensile test was performed at room temperature on a universal testing machine (SHIMADZU) with 50 N load cells. Samples of rectangular shape were used with the width of 10 mm, length of 40 mm, and height of 1 mm were used. The samples were placed between two platforms, with both ends gripped by the machine grips: upper in the moving grips and lower in the fixed grips. The rate of stretching was constantly 60 mm/min. The mechanical parameters were calculated from the stress-strain curve.

With a similar method, properties of adhesion and toughness were also measured,

which was called as the tensile-adhesion test. The tensile-adhesion test was performed at room temperature on a universal testing machine (SHIMADZU) with 50 N load cells. To measure the property of adhesion, samples with the size of 37 mm in width, 38 mm in length, and 1 mm in height were used. The samples were pre-settled onto a flat horizontal square platform. The platform had a vertical handle for the testing machine to grip. As a whole, the device was T-shaped. The samples were placed between two platforms, with both handles gripped by the machine grips: upper in the moving grips and lower in the fixed grips. The rate of stretching was constantly 60 mm/min. The related parameters were calculated from the stress-strain curve.

*Rolling Locomotion*

Rolling locomotion was achieved through a clockwise rotating magnetic field. After the robot was settled into the three-dimensional magnetic field generated by the six-coiled setup, a rotating ***B*** of 10 mT was applied in x-y plane, and the field rotated at a frequency of 1 Hz. The phase shift $\beta$ was set as 45 degrees.

1. Comparison of Rolling on Different Material Bases

Different material bases on which robots rolled over were substituted. Pieces of silicone rubber, polyurethane, and polytetrafluoroethylene were separately cut off with a size of 5 cm in width, 20 cm in length. Robots with different concentrations of porous silica were settled onto each of the surfaces and rolled within the rotating ***B***.

2. Comparison of Rolling on Slope of Different Inclinations

Since the material bases (silicone rubber, polyurethane, and polytetrafluoroethylene) of the path were soft, two wooden plates were pieced together to serve as the backbone of the slope. Patches of silicone rubber, polyurethane, and polytetrafluoroethylene were then glued to the backbones. The backbone was stuck up on one side to the top area with the other side at the ground, so that the angle formed between the plates and the ground matched with the indicated angles from the angle gauge. The rolling conditions of robots with different porous silica concentrations were measured at specific angles: 30°, 45°, and 60°. Robots were put onto the horizontal section of the material base, starting at the same location each time.

3. Application of rolling: magnetically-controlled trolley

A plastic trolley (L=3.5 cm, w=2.5 cm) was used as the basic backbone for the construction of the magnetically-controlled trolley. The original synthetic elastomer on the tire was peeled off to an extent that the circumference of the tire backbone was at the same length as the robot (8.9 mm). The magnetized microrobot was then wrapped around the backbone of the tire. The wrapping start point of tires in parallel were at the same position to ensure uniform movement. After substitution for all the four tires, the trolley was put into a rotating magnetic field $B$ in x-y plane for testing. While each trolley had four identical tires, different trolleys had tires composed of robots with different porous silica concentrations. Their motions on different material bases at different slopes were tested. Furthermore, in order to demonstrate its practical significance in cargo delivery, weights were added on the trolley to test the maximum weight that each trolley of specific porous silica concentration could carry.

*Walking Locomotion*

The walking locomotion was involved with a continuous change of magnetic field, for which each walking period was divided into three phases. As continued walking required repeated walking periods, a function was coded to apply a constantly changing field. (See supplementary material for detailed explanations.)

The phase shift $\beta$ was set as 45 degrees.

1. Comparison of Walking on Different Material Bases

   Setup same as that of rolling locomotion.

2. Comparison of Walking on Slope of Different Inclinations

   Setup same as that of rolling locomotion.

3. Comparison of Resistance to Interference of Air Flow in Robots of Different Porous Silica Concentration

   Robots were placed on the silicone rubber base with the same starting position each time. The airflow generator was placed at the side that the robot moved forward to. The airflow rate was set to different levels. The rate started off with the first level and changed to the second level when robots moved to the center. The subsequent paths of walking movements were observed and compared for robots that had different porous silica concentrations.

4. Comparison of Resistance to Interference of Water Flow in Robots of Different Porous Silica Concentration

   Water flow was generated on the moving path of the robot. A funnel was fixed at a certain height to ensure constant outflow that resulted from the uniform atmosphere pressure, and a plastic tube was connected to the outlet of the funnel. The outflow rate could be calculated by Bernoulli's equation. The outlet of the tube was right on the movement paths, which was on the silicone rubber base fitted into flat glassware capable of holding the water. The flat glassware was horizontally tilted to a mild angle so that the water flowed in a given direction.

   The robot started to walk first with the changing magnetic field. While the robots entered a relatively stable mode of walking, water was added to the funnel to produce water outflow that encountered the walking robot. The subsequent change in paths and status of movements were observed and compared for robots that had different porous silica concentrations.

*Jumping Locomotion*

There were two kinds of jumping locomotion, vertical jumping, and horizontal jumping. In both kinds of jumping, applied magnetic field was a constant field along z axis, and the magnitude varied in different attempts. For vertical jumping, the phase shift $\beta$ was set as -90 degrees. For the horizontal jumping, the phase shift $\beta$ was set as 0 degrees.

Due to experimental uncertainty, it was difficult to quantify the relationship between porous silica doping concentration and vertical jumping height. However, it was still possible to draw a comparison of height and distance reached by horizontal jumping of robots with different porous silica concentrations.

To compare the height and distance in horizontal jumping by microrobots of different porous silicon concentrations, one patch of silicone rubber and two patches of polyurethane were used, with each patch at the thickness of 0.3 cm. The material bases were placed side by side: the silicone rubber on the left and the two pieces of polyurethane overlapped. The robots jumped from the silicone rubber (h=0.3 cm) to the two overlapped polyurethane layer (h=0.6 cm in total).

In addition, a scenario of lower jumping height required was also tested in case that the prior one did not reveal differences. In this scenario, one patch of silicone rubber (h=0.3 cm), one patch of wooden plate (h=0.1cm) and one patch of polyurethane (h=0.3 cm) were used. The silicone rubber was again used as the starting plane (h=0.3 cm), while the final platform was changed to an overlapping of the wooden plate and the polyurethane layers (h=0.4 cm in total).

*Crawling Locomotion*

For crawling locomotion, the microrobot was put into a fine glass tube initially. Magnetic field **B** was adjusted to a magnitude of 10 mT, rotating in x-y plane with a frequency of 10-25 Hz. The phase shift $\beta$ was set as 45 degrees. The tube that microrobots crawled through had a diameter of 1.2 mm. During the locomotion, the microrobot moved from one side of the tube to the other.

*Drifting Locomotion*

Drifting had the same applied magnetic field as walking in essence, while it moved on the water surface due to the strong surface tension of water molecules. The magnetic field setup in drifting was the same as that in walking.

The application of drifting locomotion was a magnetically-controlled boat. Two wooden rods composed the backbone of the boat, supporting the cargo-holding plate by fixation and attaching to drifting robots that served as motors of the boat. Four robots were involved, and the central section of each robot was fixed to one end of the rod. The rest of the robot body was free for movement, allowing uniform drifting motion that propelled the boat.

*In Vivo Simulation*

The stomach model was made of silicone rubber, while the blood vessel was made of plastics. Since body fluid distributed randomly over the organs in human body, certain amount of water was supplied randomly onto the surface of both models. Also, massagers were used to simulate the peristalsis of the inner wall of muscles. Microrobots performed multimodal locomotion in such environment with *in vivo* simulation.

*Stirring Locomotion for Biocatalysis (GOx-HRP Reaction)*

1. Homogeneous Biocatalysis

The enzyme solution was prepared with 9 mg of GOx, 9 mg of HRP, and 30 mL of PBS solution. The glucose solution of 40 mmol/L (3.603 g/L) and ABTS solution of 1.064 mmol/L (0.292 g/L) were prepared with a volume ratio of 1:1. Then 20 μL enzyme solution was added into the mixture of glucose and ABTS solution. The robot was added and the stirring motion started. The essence of stirring locomotion was rolling. Therefore, a rotating magnetic field was applied. In order to achieve more adequate mixing, the microrobot was set to rotate in x-y, x-z, y-z plane respectively. The progress of this reaction was monitored through observation of the color dispersion - in comparison to another identical mixture without the robot.

For the simulation in the microreactor, multimodal locomotion was achieved. The robot walked on the left platform and then rolled into the solution. 20 μL enzyme solution was added. Then the robot began to stir on x-z plane. After the reaction, the robot rolled out of the reactor.

2. Enzyme-carrying Catalysis

20 mg of GOx and 20 mg of HRP were added into the 17.6 g material in making the 1 mm thick film that made up the robot body. Robots were cut off and magnetized. The glucose solution of 40 mmol/L (3.603 g/L) and ABTS solution of 1.064 mmol/L (0.292 g/L) were prepared with a volume ratio of 1:1. The robot was then added and the stirring motion started. The progress of this reaction was monitored through observation of the color dispersion—in comparison to another enzyme-added ABTS-glucose mixture without the robot.

Similarly, the enzyme-carrying robot was used for the simulation in the microreactor.

To explore the repeatability of catalysis in a single enzyme-carrying robot, one GOx-HRP containing robot was used for several times. After each usage, the robot was rubbed with abrasive paper to make a fresh surface. By applying a used GOx-HRP containing robot to the reaction several times and each time comparing to the enzyme-added reaction without the robot, the repeatability of catalysis in a single robot could be known.

3. GOx-HRP Cascade Reaction

40 mg of GOx was added into the 17.6 g material in making the 1 mm thick film that made up robot bodies, and 40 mg of HRP into another 17.6g material. Two kinds of robots were separately cut off from the films and magnetized. By controlling the locomotion of two robots sequentially, the step of cascade reaction could be controlled. Initially, the GOx-containing (40 mg of GOx) robot rolled into the microreactor. Then the stirring locomotion started. After 80 seconds of reaction, the GOx-containing robot rolled over the slope base and exited the environment. This completed the first step of the GOx-HRP cascade reaction, and the color of the solution did not change. Then, the HRP-containing robot rolled into the container. Stirring locomotion started. After 60 seconds of reaction, the HRP-containing robot rolled out of the container. This completed the second step of the GOx-HRP cascade reaction, and the color of solution changed to blue.

4. Heterogeneous Biocatalysis

The glucose solution of 40 mmol/L (3.603 g/L) and ABTS solution of 1.064 mmol/L (0.292 g/L) were prepared with a volume ratio of 1:1. 10 mL of glycol was added into the mixture of glucose and ABTS. The solvent was namely 50% glycol (aq). Then, the microrobot was gently put into the liquid. The enzyme solution was prepared with 9 mg of GOx, 9 mg of HRP, and 30 mL of PBS solution. 100 μL of the solution was gently dropped into the glycol mixture. A phase interface appeared between the enzyme solution and the glycol mixture. The magnetic field was turned on to start the stirring locomotion. The phase interface greatly hindered the reaction, but stirring locomotion broke the interface and made the reaction to proceed quickly. Similarly, the enzyme-carrying robot was used for the simulation in the microreactor.

*Thrombus Elimination*

The crawling robots had the potential to eliminate thrombus, as shown in the simulation. The gel-like and powder-like substances were put into the tube as two different representatives of thrombus, obstructing the forward crawling movement in the tube. With a rotating magnetic field same as that in crawling locomotion, the crawling robot scraped off the thrombus with its undulating movement.

**Supporting texts**

**Text S1. General Theory for the Soft Robot Locomotion Controlled by Magnetic Field**

Here, we first give the definition of ***m***. ***m*** meant the magnetic moment per unit volume, which was induced by input magnetic field ***B₀***.

$$\boldsymbol{m} = \frac{d\boldsymbol{M}_t}{dV}$$

(Equation S1)

***M_t*** was the total magnetization, which was also known as the net magnetic moment. ***m*** described the distribution of magnetization direction, and due to the magnetization process mentioned above, the initial ***m₀*** of the unwrapped soft robot could be described as:

$$\boldsymbol{m_{0(s)}} = [m_{0x}, m_{0y}, 0]^T = m \begin{pmatrix} \cos(\omega s + \beta) \\ \sin(\omega s + \beta) \\ 0 \end{pmatrix}$$

(Equation S2)

The variables *m₀ₓ* and *m₀ᵧ* represented the components of ***m*** along the x-axis and y-axis, respectively, while *m* and *ω* represented the magnitude and spatial angular frequency of ***m₀₍ₛ₎***, respectively. Define *L* as the length of robot body, then $\omega = 2\pi / L$.

Give an external magnetic field ***B***, which was described as $B[B_x, B_y, B_z]^T$.

First, the whole robot was taken as the research object. Then,

$$\boldsymbol{M_{t0}} = \int \boldsymbol{m} \, dV = \int_0^L \boldsymbol{m_{0(s)}} A \, ds$$

(Equation S3)

*A* was cross-section area, $A = wh$. Therefore, the initial bending moment (Mw0) of the robot as a whole came from the interaction of ***M_t*** and ***B***:

$$\boldsymbol{M_{w0}} = \boldsymbol{M_{t0}} \times \boldsymbol{B}$$

(Equation S4)

Through (Eq 4), it was known that when ***M_{t0}*** and ***B*** were not parallel, under the action of ***B***,

the robot underwent rigid rotation as a whole until $M_{t0}$ and $B$ became parallel. Before deformation, $M_{t0} = \int_0^L m_{0(s)} A ds = 0$, which meant it never underwent rigid rotation.

However, after deformation, rigid rotation may take place because when the deformation reached equilibrium after reacting with $B$, $m$ was no longer the direction before deformation, but rotated by the angle $\theta$ around the z axis. Here, $m_{(s)}$ was used to represent $m$ after deformation. In order to simplify the model, the deflection of soft robot would be analyzed in depth, ignoring other minor factors. Therefore, only the magnetic field in one plane need to be analyzed, for magnetic field vertical to the plane would affect the soft robot considering bending in the given $m$. Thus, the basic analysis could be simplified that the magnetic field was on xy plane, and $B_z = 0$.

According to the analysis above, $m_{(s)}$ could be described using the standard z-axis rotational matrix $R$

$$R = \begin{pmatrix} \cos\theta & -\sin\theta & 0 \\ \sin\theta & \cos\theta & 0 \\ 0 & 0 & 1 \end{pmatrix}$$

(Equation S5)

$$m_{(s)} = R m_{0(s)}$$

(Equation S6)

Second, an infinitesimal element on the robot was considered, and according to the geometry, the relationship between curvature ($1/\rho$), rotation angle ($\theta$) and deformation along the body (d$s$) could be obtained (Figure S1)

$$\frac{1}{\rho} = \frac{\partial \theta_{(s)}}{\partial s}$$

(Equation S7)

And the relationship between curvature ($1/\rho$) and bending moment along z axis ($M_z$) was also expressed as

$$\frac{1}{\rho} = \frac{M_z}{EI_z}$$

(Equation S8)

Therefore,

$$M_z = EI_z \frac{\partial \theta_{(s)}}{\partial s}$$

(Equation S9)

Under quasi-static conditions, the moments of the micro element were balanced, so the balance equation could be expressed as:

$$M_z = (M_z + dM_z) + \Delta M_z$$

(Equation S10)

Here, $M_z$ and $M_z+dM_z$ represented the moments experienced by the two ends of the element, respectively, while $\Delta M_z$ was the moments experienced by the middle part of the element. So (Eq 10) could be simplified as:

$$\Delta M_z = -EI_z \frac{\partial^2 \theta_{(s)}}{\partial s^2} ds$$

(Equation S11)

Introduce a new physical quantity $\tau_z$, which was defined as the magnetic torque per unit volume along z axis:

$$\boldsymbol{\tau_z} = R\boldsymbol{m_0} \times \boldsymbol{B}$$

(Equation S12)

Therefore, Equation S11 could be expressed as:

$$\tau_z A = -EI_z \frac{\partial^2 \theta_{(s)}}{\partial s^2}$$

(Equation S13)

Which was the basic equation for quasi-static analysis. Here, $\tau_z$ was the magnitude of $\boldsymbol{\tau_z}$, $\tau_z = [001]R\boldsymbol{m_0} \times \boldsymbol{B}$. $I_z$ was the moment of inertia of z axis, and it was easily calculated that

$I_z = \frac{wh^3}{12}$.

So far, the relationship between robot deformation and force has been obtained, including overall force and micro-element force. The following discussions could be divided into two cases, small deformation and large deformation.

**Small deflection**

When the deformation was small, some approximations were reasonable. First, $s \approx x$. And according to geometry, $tan\theta = \frac{dy}{dx} \approx \theta$. Second, since the deformation was small, the change in the direction of $\boldsymbol{m}$ could be neglected, which meant $R = \begin{pmatrix} 1 & 0 & 0 \\ 0 & 1 & 0 \\ 0 & 0 & 1 \end{pmatrix}$.

Therefore, Equation S13 could be simplified as:

$$\tau_z A ds = [001]\boldsymbol{m} \times \boldsymbol{B} A dx = -EI_z \frac{\partial^3 y}{\partial x^3}$$

(Equation S14)

Start with the simplest case, when $\boldsymbol{B}$ was parallel to $[cos\beta, sin\beta, 0]^T$. Namely, $\boldsymbol{B} = B[cos\beta, sin\beta, 0]^T$. Then solve (Eq 14) to obtain the relationship between y and x,

$$y_{(x)} = \frac{mBA}{EI_z \omega^3} cos\omega x + C_1 x^2 + C_2 x + C_3$$

(Equation S15)

$C_1$, $C_2$, $C_3$ were constants. Considering the special circumstance, when $\boldsymbol{B} = 0$, the robot would stay flat for all x, which meant $y_{(x)} = 0$ for all x. So $C_1$, $C_2$ and $C_3$ had to be zero.

Therefore,

$$y_{(x)} = \frac{mBA}{EI_z \omega^3} cos\omega x, 0 < x < L$$

(Equation S16)

Which was the basic equation for small deflection analysis.

After deformation, $M_t$ was not necessarily parallel to $B$. Now, $R = \begin{pmatrix} \cos\theta & -\sin\theta & 0 \\ \sin\theta & \cos\theta & 0 \\ 0 & 0 & 1 \end{pmatrix}$ and $\boldsymbol{m}_{(s)} = R\boldsymbol{m}_{0(s)}$, $\theta = \frac{dy}{dx} = \frac{-mBA}{EI_z\omega^2}\sin\omega x$.

Therefore, $R$ could be described as:

$$R = \begin{pmatrix} 1 & \frac{mBA}{EI_z\omega^2}\sin\omega x & 0 \\ \frac{-mBA}{EI_z\omega^2}\sin\omega x & 1 & 0 \\ 0 & 0 & 1 \end{pmatrix}$$

(Equation S17)

Thus,

$$\boldsymbol{M_t} = \int_0^L R\boldsymbol{m}_{0(x)} A dx$$

$$M_{tx} = Am\int_0^L \cos(\omega x + \beta) + \frac{mBA}{EI_z\omega^2}\sin\omega x \sin(\omega x + \beta)\, dx = \frac{m^2 B\pi A^2}{EI_z\omega^3}\cos\beta$$

$$M_{ty} = Am\int_0^L \sin(\omega x + \beta) - \frac{mBA}{EI_z\omega^2}\sin\omega x \cos(\omega x + \beta)\, dx = \frac{m^2 B\pi A^2}{EI_z\omega^3}\sin\beta$$

$$M_{tz} = 0$$

(Equation S18)

Therefore, it was concluded that $M_t \mathbin{\!/\mkern-5mu/\!} B$ even after small deformation.

When $B$ was perpendicular to $[\cos\beta, \sin\beta, 0]^T$, $\boldsymbol{B} = B[-\sin\beta, \cos\beta, 0]^T$, similarly, the exact relationship between y and x through the special circumstance (B = 0) could also be obtained

$$y_{(x)} = \frac{mBA}{EI_z\omega^3}\sin\omega x, 0 < x < L$$

(Equation S19)

And it was proved that $M_t \mathbin{\!/\mkern-5mu/\!} B$ also in this situation.

In conclusion, when $B$ was parallel and perpendicular to $[\cos\beta, \sin\beta, 0]^T$, the relationship between y and x exhibited the form of simple cosine and sine functions, respectively. When $B$ was in any direction, it could be decomposed into $[cos\beta, sin\beta, 0]^T$ and $[-sin\beta, cos\beta, 0]^T$ directions according to Fourier expansion, then the deformation and movement of the robot could be regarded as the superposition of sine and cosine function in these two directions. Since this was a complete set of orthogonal bases, then such decomposition could be achieved. And it was also proved that, when **B** was small, **$M_t$** // **B** was always correct, which meant the robot would not undergo rigid rotation under small deformation circumstances.

**Large deflection**

However, when the deformation was large, the assumptions in 2.2.1 no longer held true. Therefore, the relationship between $\theta$ and $s$ could only be obtained by solving Equation S13

$$B_y \cos(\omega s + \beta + \theta) - B_x \sin(\omega s + \beta + \theta) = -\frac{EI_z}{Am}\frac{\partial^2 \theta_{(s)}}{\partial s^2}$$

(Equation S20)

Using boundary conditions $M_{edge} = 0$

$$\frac{\partial \theta_{(s)}}{\partial s}|_{s=0} = \frac{\partial \theta_{(s)}}{\partial s}|_{s=L} = 0$$

(Equation S21)

Obviously, this was a second order differential equation with boundary conditions. Analytic solutions could not be achieved, but it was still possible to get its numerical solutions.

According to the dimensions of the soft robot, the parameters were listed below. $L$ = 3.7 mm, $\beta$ = $\pi/4$, $E$ = 8.5×10$^4$ Pa, $I/A$ = 2.852×10$^{-9}$, $EI/A$ = 2.852×10$^{-5}$, and $|m|$ = 77500 A/m.

Therefore, according to the international system of units, the equation could be expressed as

$$B_y \cos(1698.16s + \frac{\pi}{4} + \theta) - B_x \sin(1698.16s + \frac{\pi}{4} + \theta) = -3.91 \times 10^{-9} \frac{\partial^2 \theta_{(s)}}{\partial s^2}$$

(Equation S22)

**Text S2. A more detailed analysis of rolling locomotion**

The external magnetic field **B** was described as $B = [B_x\ B_y\ B_z]^T$. **B** was a function of time $t$. In rolling locomotion, the sequence of **B** could be described as:

$$B(t) = B \begin{pmatrix} \cos(2\pi ft) \\ \sin(2\pi ft) \\ 0 \end{pmatrix}$$

(Equation S23)

The net magnetic moment, $M_{net}$, is a function of the microrobot rotational displacement $\phi_R$, and it could be described as

$$M_{net}(\phi_R) = M_{net} \begin{pmatrix} \cos(\phi_R) \\ 0 \\ \sin(\phi_R) \end{pmatrix}$$

(Equation S24)

When **B** started, a rigid-body torque, $\tau_{Roll}$, was applied to the robot. The robot began to roll. It could be expressed as

$$\tau_{Roll} = [0\ 0\ 1](M_{net} \times B) = M_{net} B \sin(2\pi ft - \phi_R)$$

(Equation S25)

The magnitude of $\tau_{Roll}$ depended on the $M_{net}$ and **B**, and the angle between them. Under experimental conditions, they were unchanged among each experimental group.

The translational equation of microrobot locomotion was

$$F - C_T V_{roll} = M_{Rob} \dot{V}_{roll}$$

(Equation S26)

Where $C_T$ represented the translational damping. $F$ was a pushing force from the substrate, which mainly consisted of frictional force and adhesive force. Under experimental conditions, $C_T$ and $M_{Rob}$ were unchanged among each experimental group. When equilibrium was reached, that was, the robot rolled at a uniform speed, both sides of the equation were zero. And $V_{roll}$ was decided by the equation

$$V_{roll} = 2\pi r_{eff} f$$

(Equation S27)

Where $r_{eff}$ represented the radius of the rolling robot, decided by robot size, which were the same among groups. Thus, $F$ needed to reach a certain value, and since the above analysis did not consider environmental resistance, the actual $F$ needed to be greater, based on resistance. Adhesive force became especially important in case of small frictional force or large resistance.

**Text S3. A more detailed analysis of vertical jumping locomotion**

When we set $\beta = -90° = -\frac{\pi}{2}$ and $B_x = 0$, the vertical jumping situation could be obtained. Therefore, the simplified equation could be expressed as

$$B_y \cos\left(1698.16s - \frac{\pi}{2} + \theta\right) = -3.91 \times 10^{-9} \frac{\partial^2 \theta_{(s)}}{\partial s^2}$$

(Equation S28)

We tried different $B_y$ and got the corresponding deflection curves (Figure 7a), from which we could get information about what magnitude of the magnetic field could make the soft robot jump up from the substrate.

According to the results above, we could analyze how the magnetic field could make the robot jump up from the substrate.

With $\beta = -90°$, the magnetization profile was illustrated (Figure S3). In this situation, a magnetic field $\boldsymbol{B}$ with $y$ direction was provided. Since $\boldsymbol{M} = \boldsymbol{m} \times \boldsymbol{B}$, initially, the left part of the soft robot body had the tendency of counterclockwise rotation, while the right part of the robot body had the tendency of clockwise rotation. This made the middle part of the robot body go up, while two ends kept in touch with the substrate, forming a small deflection convex curve. If the small deflection convex curve accorded with the final stable shape, then the robot body would keep this shape on the substrate. However, if the small deflection convex curve did not fit with the final stable shape, then the robot body might undergo other kinds of deformation. The robot could jump up when the deformation must occur but could not be accomplished relying on the support of substrate.

When $B_y = 1$ mT/3 mT/5 mT, the final stable shape was just a small deflection convex curve. Therefore, the robot body would just keep this stable shape on the substrate, never jumping up.

In contrast, when $B_y$ = 6 mT, the performance of solutions changed greatly, with the shape of deflection curve changing from convex curve to concave curve. Therefore, the small deflection convex curve formed in initial deformation was not stable, for the final state should be a concave curve with middle deflection. Thus, the robot body had a tendency to change to the final state. However, with the middle part of robot body going up, the convex curved body could not turn to concave curve easily in the deforming process. Then it would continue deformation, accumulating elastic strain energy and kinetic energy although it would not reach the stable point, and finally jumped up at one certain point so as to release kinetic energy and turned it to gravitational potential energy, achieving the high jump.

Nevertheless, the robot did not actually jump up when $B_y$ = 6 mT, due to the fact that $B_y$ = 6 mT was just near the critical point where the small deflection convex curve formed in initial deformation turned to be unstable, so the driving force for jumping seemed too small. Since there were other impediments for the robot body to jump up, such as the frictional energy loss, such small impetus would not cause the locomotion of vertical jumping. Instead, the initial $B_y$ = 6 mT just counterbalanced the frictional energy loss (contact friction with the substrate, air drag, and viscoelastic losses within the soft robot), making the locomotion of jumping possible in stronger magnetic field.

When $B_y$ was from 6 mT to 9 mT, the degree of bending for stable deflection curve enhanced, which did not change the tendency for the robot to jump up, while enhancing the impetus for the robot body to jump up. When the $B_y$ was larger than 9 mT, the stable deflection curve did not change apparently. However, the larger magnitude of magnetic field still provided more impetus for the locomotion of jumping up.

In our experiment, the lowest $B_y$ for the soft robot body to jump up was 20.1 mT (Figure 32, Video S3), which roughly equaled to the result of 18.9 mT in previous studies. The magnitude of 20.1 mT accorded with our theoretical analysis above that 6 mT of magnetic field counterbalanced the frictional energy loss and provided the basic tendency for the robot to jump up, while a higher magnitude of magnetic field provided more impetus for the locomotion of jumping up. In our experiment, approximately 14.1 mT of magnetic field was used to provide additional impetus (could be seen as kinetic energy) for jumping up.

Next, the straight jumping process was further analyzed from the perspective of energy. We defined the initial state as the state where the robot body lay flat and still on the substrate, and the jumping critical state as the state where the robot body was ready to jump up with a little bit more magnitude of magnetic field.

Based on the principle of energy conservation, between the two states, the soft robot gained energy from the work done by the magnetic torque $W$. This energy was then re-distributed into three components: the change in strain energy $\Delta S$, and kinetic energy $\Delta K$, and the frictional losses $Q$, during the jumping process. Namely, this implied that

$$W - \Delta S - Q = \Delta K$$

(Equation S29)

For our experiment, 6 mT of magnetic field provided the basic tendency for the robot to jump up, compensating for the energy loss, while approximately 14.1 mT of magnetic field was used to provide additional impetus for jumping up. We could simplify the process by making an assumption that the magnetic torque $W$ caused by the initial 6 mT of magnetic field all turned into frictional losses $Q$, which became the basic tendency for the robot to jump up. This approximation might seem abrupt, but in fact it was reasonable. Initially, the energy input by magnetic field mainly dissipated by quasi-static frictional energy losses, in which process the robot body adjusted itself to the posture of jumping permission in the quasi-static process, gaining basic tendency to jump up. Just as a block that slided uniformly down a slope, all the gravitational potential energy turned into heat. Also, it was assumed that the magnetic torque $W$ caused by the sequential 14.1 mT of magnetic field turned into three parts—strain energy $\Delta S$, kinetic energy $\Delta K$, and frictional losses $Q$. Due to the fact that additional magnetic field caused the motion of one direction which made the robot finally jump instead of quasi-static process, frictional losses $Q$ still existed but did not belong to the overwhelming factor, so we assumed the frictional losses $Q$ equaled to half of the magnetic torque $W$ in this part. Just as an object that was accelerated by a conveyor belt, half of the energy provided became frictional losses, the other half of energy made the object accelerate.

According to our approximation, (6+(14.1/2))/20.1=64.93% of the magnetic torque $W$ would turn into frictional losses $Q$, while (14.1/2)/20.1=35.07% of the magnetic torque $W$ would be transformed into strain energy $\Delta S$ and kinetic energy $\Delta K$. Thus, the equation above could be simplified as

$$0.3507W - \Delta S = \Delta K$$

(Equation S30)

For an infinitesimal element $ds$,

$$M = \frac{EI}{ds} d\theta$$

(Equation S31)

Namely, in the infinitesimal element $ds$, the magnetic torque $W$ was proportional to the angle $d\theta$. Compared to the elastic potential energy of a spring, and assume that the $\theta_i$ at initial state equaled to zero (on a flat substrate), then

$$dM = \frac{1}{2}Md\theta_f = \frac{1}{2}EI\left(\frac{d\theta_f}{ds}\right)^2 ds$$

(Equation S32)

The strain energy $\Delta S$ equaled to the work needed to bend the whole body, thus,

$$\Delta S = \int_0^L dM = \frac{1}{2}\int_0^L EI\left(\frac{d\theta_f}{ds}\right)^2 ds$$

(Equation S33)

On the other hand, the magnetic work done $W$ could be expressed as

$$W = \int_0^L F ds = \int_0^L \left[\int_0^{\theta_f} \tau_z A \, d\theta\right] ds$$

Since

$$\tau_z = [001]R\boldsymbol{m_0} \times \boldsymbol{B}$$

$$R = \begin{pmatrix} \cos\theta & -\sin\theta & 0 \\ \sin\theta & \cos\theta & 0 \\ 0 & 0 & 1 \end{pmatrix}$$

Then

$$\int_0^{\theta_f} \tau_z A d\theta = \int_0^{\theta_f} \{(m_x B_y - m_y B_x)\cos\theta - (m_y B_y + m_x B_x)\sin\theta\} A d\theta$$

(Equation S35)

Since the magnetization profile and the magnetic field were independent of $\theta$, we could calculate the integration above and get the result below

$$\int_0^{\theta_f} \tau_z A d\theta = \{(m_x B_y - m_y B_x)\sin\theta_f + (m_y B_y + m_x B_x)(\cos\theta_f - 1)\} A$$

(Equation S36)

For the vertical jumping situation, $B_x = 0$, $B_y = B$, thus

$$W = -mBA \int_0^L \cos(\omega s + \theta_f) ds$$

$$-0.3507 mBA \int_0^L \cos(\omega s + \theta_f) ds - \frac{1}{2} \int_0^L EI\left(\frac{d\theta_f}{ds}\right)^2 ds = \Delta K$$

(Equation S37)

At the maximum height, we assumed that all the kinetic energy $\Delta K$ is fully converted into the gravitational potential energy. In fact, part of strain energy $\Delta S$ should also turn into gravitational potential energy, but considering the shape of soft robot body did not change a lot, the reduction of strain energy $\Delta S$ could be omitted. Therefore,

$$\Delta K = m_r g H_{max}$$

(Equation S38)

Where $m_r$ represented the mass of the robot.

From the equations above, $H_{max}$ could be obtained as

$$H_{max} = \left( 0.3507 \frac{mBA}{m_r g} \int_0^L \cos(\omega s + \theta_f) ds + \frac{EI}{2m_r g} \int_0^L \left(\frac{d\theta_f}{ds}\right)^2 ds \right)$$

(Equation S39)

Since $|m|=62000$ A/m, $A = wh = 2.775\times10^{-7}$ m², $E = 8.5\times10^4$ Pa, $I = 7.91453\times10^{-16}$ m⁴, $m_r = 1.9098\times10^{-6}$ kg, $g = 9.8015$ m/s², $B = 20.1$ mT, the equation could be simplified as

$$H_{max} = \left( 6.47897 \int_0^{0.0037} \cos(1698.16s + \theta_f) ds + 1.796940 \times 10^{-6} \int_0^{0.0037} \left(\frac{d\theta_f}{ds}\right)^2 ds \right)$$

(Equation S40)

In order to calculate the $H_{max}$ eventually, we should get $\theta_f(s)$. It was hard to get this function from theoretical analysis. Instead, we needed to use a camera to record the real process of jumping, find the jumping critical point and get the $\theta_f(s)$ from the picture of the jumping critical point. It was a complex work for it put forward high requirements for photographing technology, and experimental errors were not negligible.

However, there was another method for us to get the $\theta_f(s)$. From the analysis above, we came to a conclusion that the theoretical critical jumping point occurred when the $B_y$ was from 5 mT to 6 mT, although the jumping critical point in real experiment required the magnetic field to be stronger. 5 mT situation corresponded to the relatively stable shape, while 6 mT situation gave a totally different shape which caused reverse deformation that induced the locomotion of jumping. Therefore, we could make an assumption that the stable state of the 5 mT situation was roughly the shape of robot body at the critical jumping point. Using the solutions $\theta(s)$ given by MATLAB for stable state of the 5 mT situation as the $\theta_f(s)$, we could do the integration above so as to get $H_{max}$ by infinitesimal analysis.

According to calculation,

$$H_{max} = 12.3 \text{ mm}$$

(Equation S41)

This result seemed to be a smaller approximation compared with the $H_{max}$ given by previous research, which gave the theoretical calculation value of 25.9 mm. However, we should notice that the actual jumping height was always lower than the theoretical results, with 6.8 mm in our experiment and 4.9 mm in previous research. Therefore, in our research, the theoretical result was much closer to the real jumping height in experiment, giving a more accurate approximation. This proved the assumption that the stable state of the 5 mT situation was roughly the shape of robot body at the critical jumping point was reasonable. Although the relative error was still non-negligible, due to the fact that the measurement and calculation were so complicated for the soft robot that much approximation must be introduced, and previous research only showed higher relative error, our study has made significant improvements.

**Text S4. A more detailed analysis of crawling locomotion**

If a periodic rotating magnetic field was applied, the robot would hit the pipe wall periodically, so it was possible to move forward continuously with a cosine shape, which was an analogy to the locomotion of undulating swimming. Here, a physical model based on analogy to mechanical wave was raised to explain the essence of crawling locomotion.

According to the general theory, in a periodic rotating magnetic field, crawling microrobot would undergo small deformation, and the deflection curve could be expressed as

$$y_{(x)} = \frac{mBS}{EI_z\omega^3}cos\omega x, 0 < x < L$$

(Equation S42)

In the crawling process, at any fixed time, the microrobot showed the shape of a cosine wave function, just like the mathematic relationship given in Eq 9. According to the above fluctuation information, naturally, it was supposed that the crawling locomotion may be expressed as a mechanical wave. However, the mechanical wave itself was a form of propagation between adjacent particles of elastic medium, and the particles themselves would not move in the direction of wave propagation. And the microrobot would certainly move forward during crawling locomotion, which constructed a contradiction with typical mechanical wave. In order to solve the problem, it could be assumed that elastic medium was arranged along the length of the whole tube, and the robot body represented a whole wavelength, going forward with the hypothetical propagation of mechanical wave. Due to boundary conditions of a single microrobot, the robot body could only represent one wavelength but not many. In this way, the forward dynamics of a microrobot entity were described by utilizing the phase propagation in the mechanical wave.

Next, the mechanical wave function should be worked out for quantitative analysis. Since actuated by rotating magnetic field synchronously, the frequency of rotating magnetic field should also be the frequency of mechanical wave. In fact, for each period of rotating magnetic field, the microrobot performed two properties together. First, for each period of rotating magnetic field, the microrobot itself would go up and down for a complete cycle, just like a standing wave. Second, in each period, the microrobot would go forward and achieve phase propagation, just like a traveling wave. However, it was not a typical traveling wave for the microrobot did not go forward for an entire wavelength (which was a must for traveling wave), but only a part of it. Thus, the mechanical wave should be described as a phase-propagation-hindered traveling wave. In other words, the wave could still travel, but not so fast as a typical travelling wave.

Thus, an important question was that how the phase propagation was hindered. It should be noticed that Eq 9 described the small deflection curve for free moving, but not in a confining tube. Therefore, the amplitude $A = \frac{mBS}{EI_z\omega^3}$, which was only suitable in free propagation, became no longer feasible owing to the boundary of the glass tunnel.

Since

$$A = \frac{mBS}{EI_z\omega^3} = \frac{mBSL^3}{8EI_z\pi^3}$$

(Equation S43)

It was calculated that

$$A = 1.20 \times 10^{-3} \text{ m}$$

(Equation S44)

Thus, the free-moving amplitude should be 1.2 mm. However, the radius of glass tunnel was only 0.6 mm, which indicated that the microrobot crawling locomotion must be confined in the glass tunnel, and a free moving was impossible. Make a comparison between the traveling wave in the free environment and the phase-propagation-hindered traveling wave in the glass tube, then it was inferred that the confinement of the tube wall caused the hindrance for the phase propagation. Any glass tube with a radius less than 1.2 mm could cause this effect, and a smaller radius would enhance the hindrance and make the microrobot march slower. It would be greatly complicated to give an analytical explanation of how this confinement hindered the phase propagation, but an intuitive description could still be achieved. A direct idea was that the ratio between glass tube radius and free moving amplitude could describe the effect of phase propagation hindrance. However, such a model could not fit with experimental results well. The inner reason was that the hindrance effect of tube boundary on phase propagation could not be a linear relationship. According to the small deflection analysis above, it was concluded that the effect of global variable on deflection curve was closer to a cubic relationship. Therefore, the cubic of the aforementioned ratio could be adopted to show the phase propagation hindrance, which fitted well to experimental results.

Thus, the wave function describing crawling locomotion could be expressed as

$$y_{(x,t)} = A'cos(\omega x + 2\pi\gamma f t), 0 < x < L$$

(Equation S45)

In this equation, A' = 0.6 mm, which was the radius of the glass tunnel. And $\gamma$ was a damping coefficient that implied how the propagation of wave was influenced by the tube wall, $f$ was the wave frequency, which equaled to the frequency of the applied rotating magnetic field. The influence of gravity was negligible in this case. The coefficient $\gamma$ could be expressed as the cubic of the ratio of bounded amplitude and free amplitude

$$\gamma = (\frac{A'}{A})^3 = 0.125$$

(Equation S46)

Since the robot body was seen as a whole period and wavelength $\lambda$ was just microrobot body length $L$, the speed of robot $v$ equaled to the wave speed $u$. Considering the phase propagation delay coefficient $\gamma$, the actual speed of microrobot could be expressed as

$$v = \gamma L f$$

Substituting $\gamma = 0.50$, $L = 3.7$ mm and $f = 15$ Hz, it could be calculated that

$$v = 6.94 \text{ mm·s}^{-1}$$

(Equation S48)

The speed of microrobot $v$ indicated how the robot crawled in a glass tunnel with a rotating magnetic field.

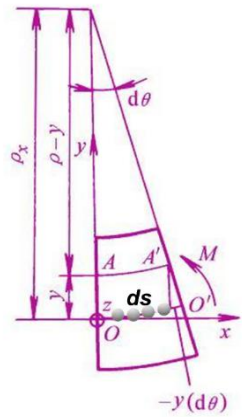

**Figure S1. An illustration of the relationship between curvature (1/$\rho$), rotation angle ($\theta$) and deformation along the body (d$s$).** What worth mention was that d$s$ was used instead of dx, for the former applied to all bending situations, but the latter only worked for small deformation.

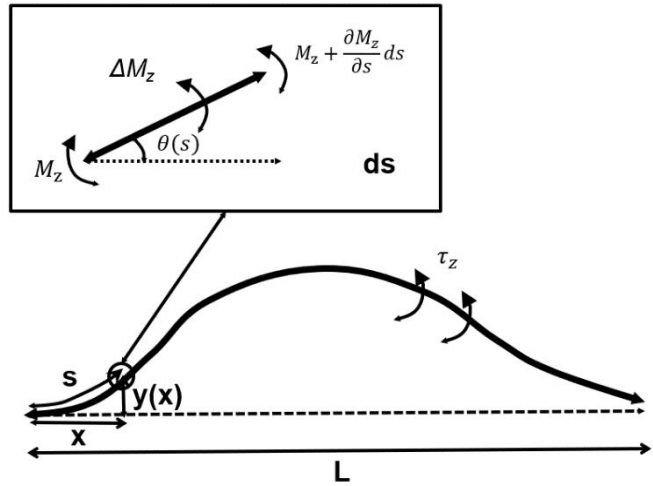

**Figure S2. Quasi-static analysis of the soft robot.** This was the foundation for locomotion analysis of the magnetic soft microrobot.

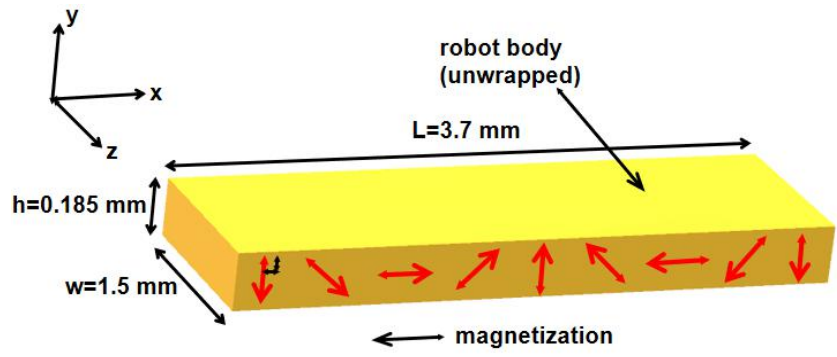

**Figure S3. The magnetization profile of the unwrapped robot body with** $\beta = -90°$**.** This was for vertical jumping. Initially, the soft robot was spread on a flat oiled paper surface with no deflection.

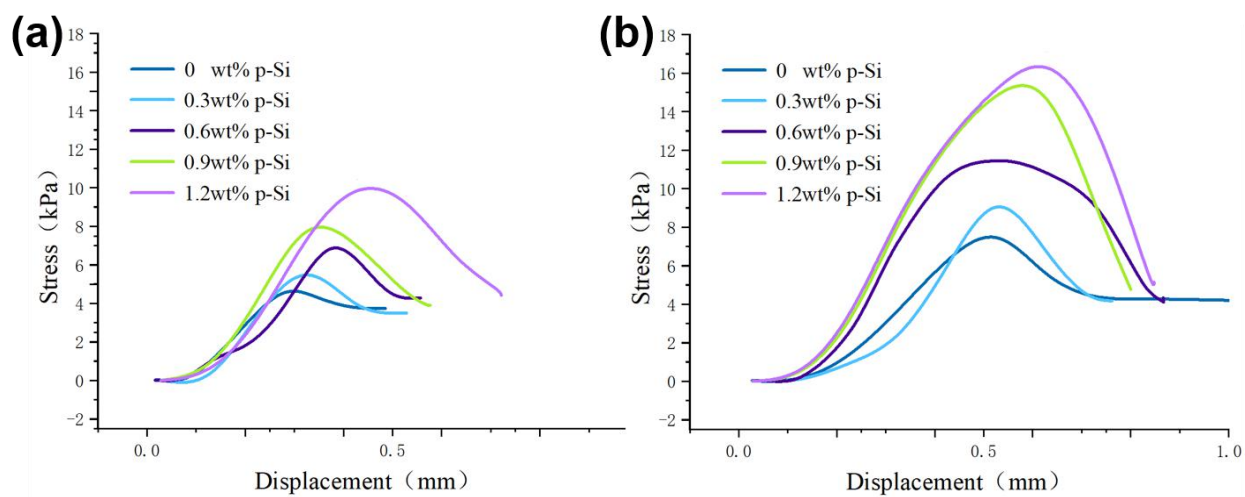

**Figure S4. Stress-strain curves for the tensile-adhesion test.** (a) For silicone rubber substrate. (b) For PTFE substrate.

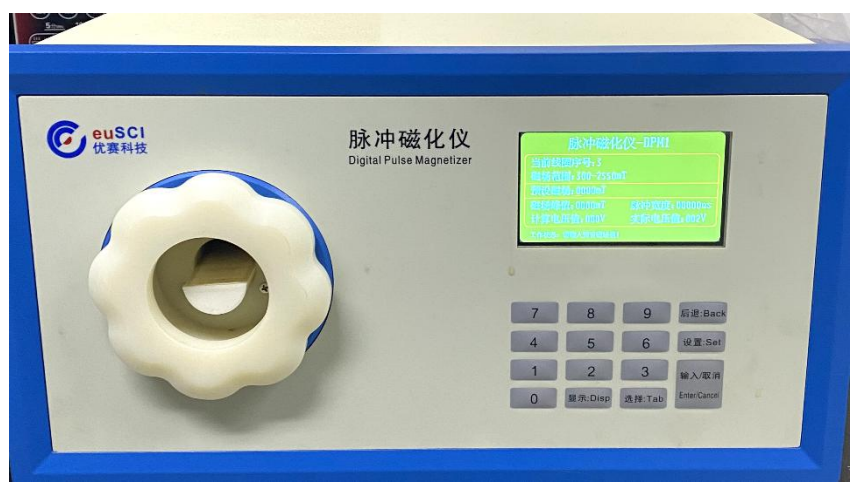

**Figure S5. The pulse magnetizer for magnetization.** The properly wrapped microrobot was placed into the pulse magnetizer (Beijing EUSCI Technology Limited) for magnetization with a strong, directional, uniform magnetic field ***B*** of 2.5 T.

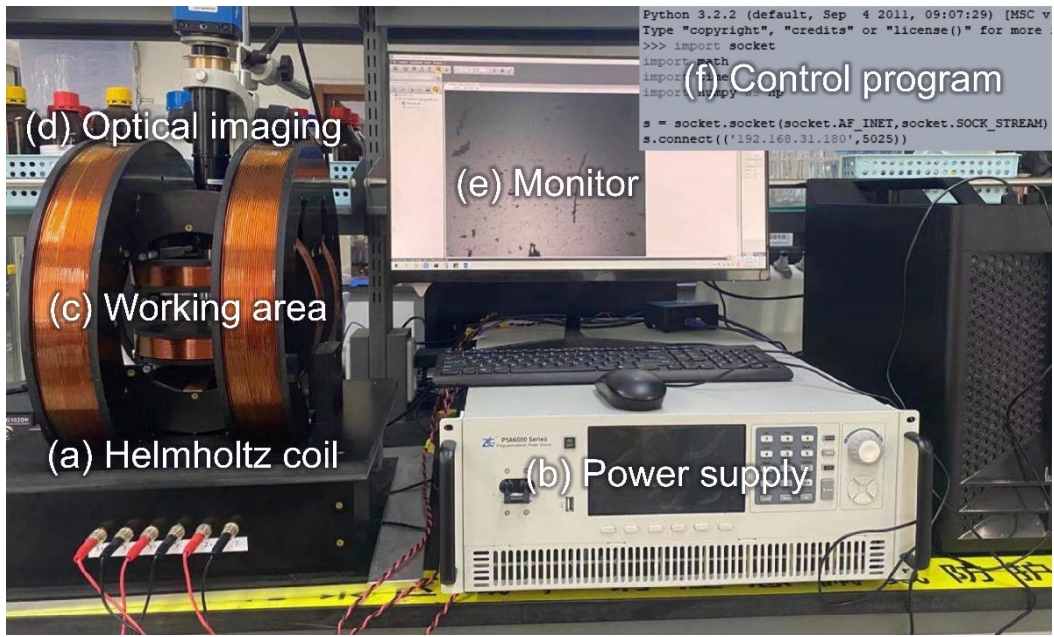

**Figure S6. The self-built three-dimensional magnetic field control system used in this study.** (a) The core device was composed of three orthogonal coils. According to the basic principle of electromagnetic induction, the magnetic field in three directions (x, y, z) could be generated. (b) The voltage of the magnetic field coil was basically regulated by a controllable voltage source (AC / DC) (Guangzhou ZHIYUAN Electronics Co., Ltd. PAS6006-3). (c) The working area was placed inside the magnetic coil, where a magnetic field of three dimensions could act simultaneously. In this research, all the locomotions to be studied were placed in the working area. (d) The optical imaging system was placed in the center of the magnetic field coil and above the working area to observe multimodal locomotion. (e) A monitor was used for the videographic display of the optical imaging system. (f) The fine adjustment of magnetic field intensity was carried out by a Python program. With the Python program, automatic control of multimodal locomotion could be achieved in the laboratory.

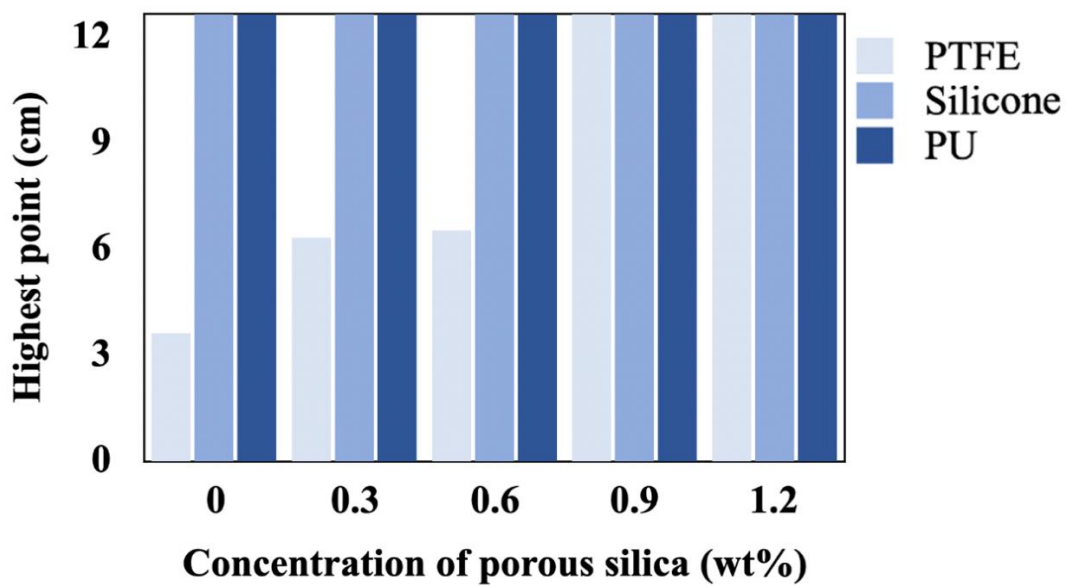

**Figure S7. Slope experiment for rolling locomotion.** A bar graph was drawn to show the highest reaching point of rolling locomotion for different doping concentrations and different substrate materials at the slope of 30 degrees.

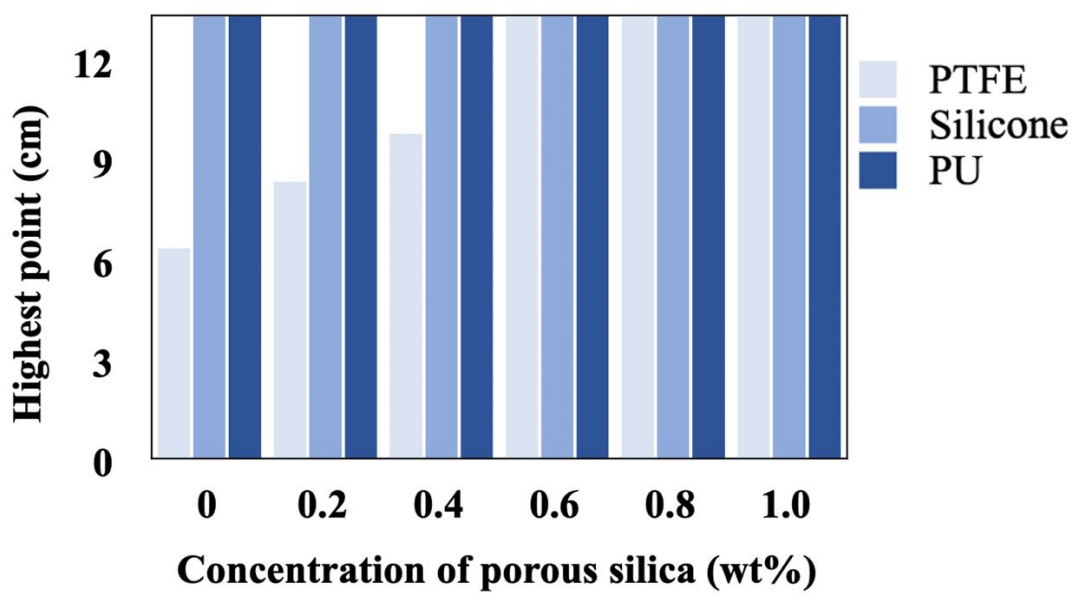

**Figure S8. Slope experiment for rolling locomotion.** A bar graph was drawn to show the highest reaching point of magnetically-controlled trolley for different doping concentrations and different substrate materials at the slope of 30 degrees.

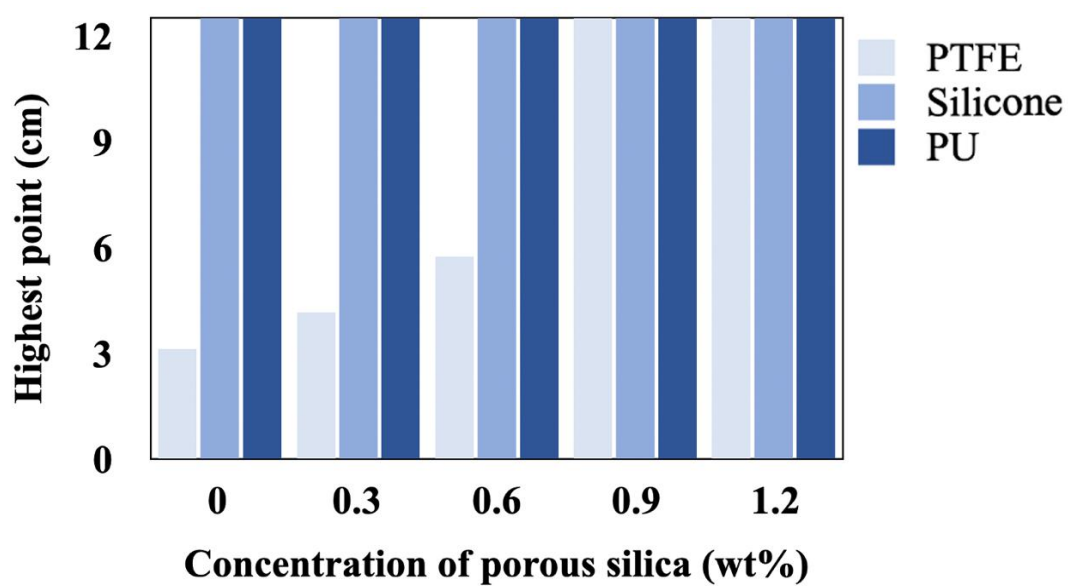

**Figure S9. Slope experiment for rolling locomotion.** A bar graph was drawn to show the highest reaching point of rolling locomotion for different doping concentrations and different substrate materials at the slope of 45 degrees.

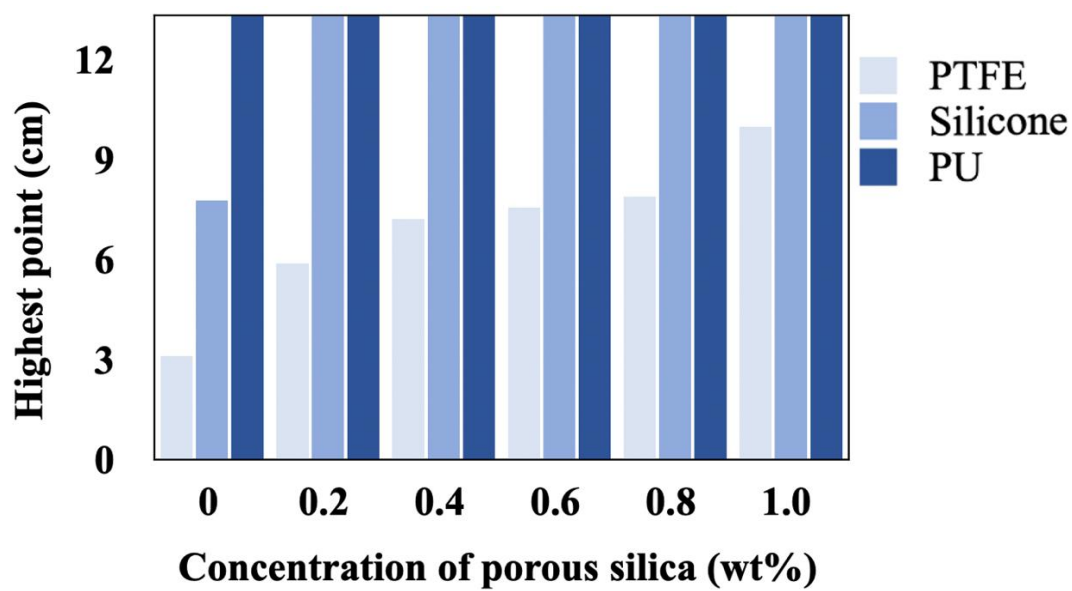

**Figure S10. Slope experiment for rolling locomotion.** A bar graph was drawn to show the highest reaching point of magnetically-controlled trolley for different doping concentrations and different substrate materials at the slope of 45 degrees.

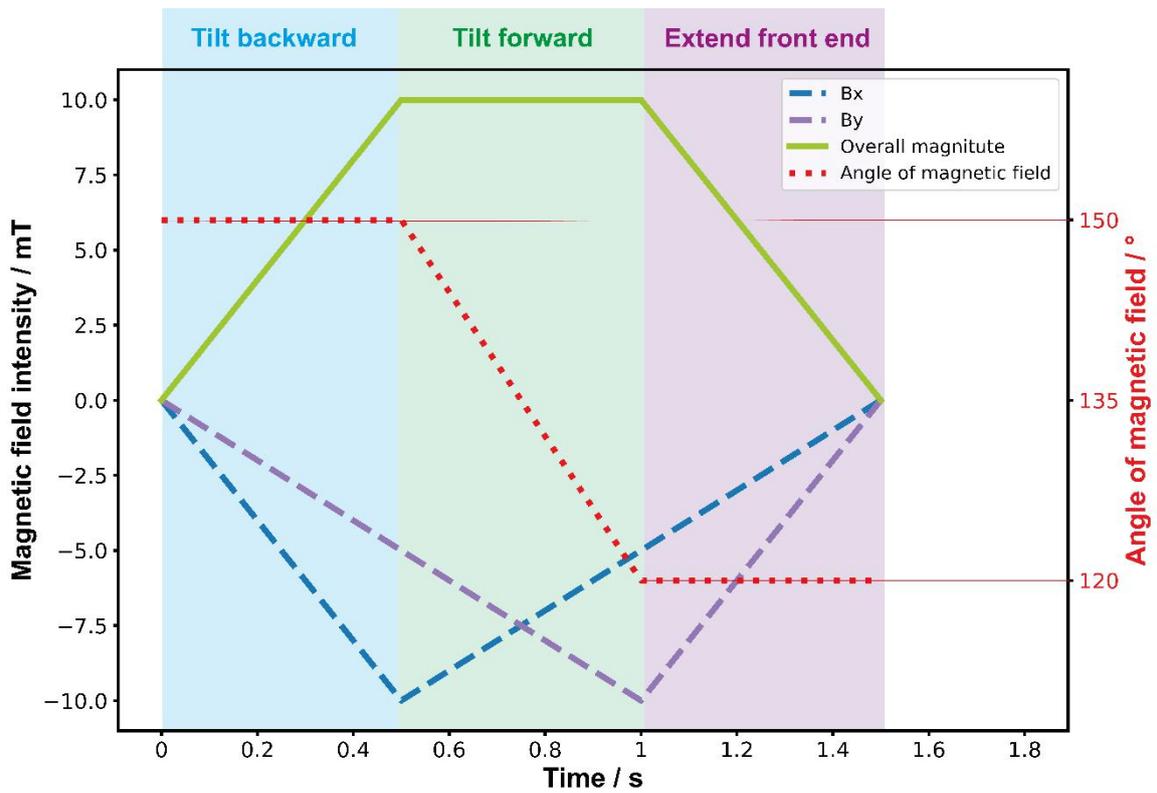

**Figure S11. The periodic magnetic field for walking locomotion.** For each period, $B_x$ and $B_y$ changed continuously, thus making the microrobot tilt backward, tilt forward and extend the front end. Therefore, the microrobot could achieve walking locomotion.